\documentstyle[12pt,a41]{article}
\newcommand{\ra}{\rightarrow}
\newcommand{\la}{\leftarrow}
\newcommand{\E}{\rm E}

\newcommand{\Li}{\mbox{Li}}
\newcommand{\SN}{\mbox{S}}
\newcommand{\Sf}{\mbox{S}_{1,2}}
\newcommand{\M}{\mbox{\rm\bf M}}
\newcommand{\N}{\mbox{\rm\bf N}}
\newcommand{\Mvec}{\mbox{\rm\bf M}}

\newcommand{\beq}{\begin{equation}}
\newcommand{\eeq}{\end{equation}}
\newcommand{\bea}{\begin{eqnarray}}
\newcommand{\eea}{\end{eqnarray}}

\newcommand{\lsim}{\raisebox{-0.07cm}{$\, \stackrel{<}{{\scriptstyle
\sim}}\, $}}

\newcommand\ds{\displaystyle}

\newcounter{lin}

\begin{document}
\begin{titlepage}

\begin{flushleft}
DESY 98--149 \hfill {\tt hep-ph/0003100} \\
March    2000                         \\
\end{flushleft}

\vspace{3cm}
\begin{center}
{\LARGE\bf Analytic Continuation of Mellin Transforms}

\vspace{3mm}
{\LARGE\bf up to two--loop Order}

\vspace{4cm}
{\large Johannes Bl\"umlein}

\vspace{2cm}
{\large\it DESY--Zeuthen}\\

\vspace{3mm}
{\large\it  Platanenallee 6, D--15735 Zeuthen, Germany}\\
%%\today

\vspace{3cm}
\end{center}
\begin{abstract}
\noindent
The analytic continuation of the Mellin transforms to complex values 
of $N$ for the basic functions $g_i(x)$ of the momentum fraction $x$ 
emerging in the quantities of massless QED and QCD up to two--loop
order, as the unpolarized and polarized splitting functions, coefficient 
functions, and hard scattering cross sections for space- and time-like 
momentum transfer are evaluated. These Mellin transforms provide the 
analytic continuations of all finite harmonic sums up to the level of 
the threefold sums of transcendentality four, where the basis--set 
${g_i(x)}$ consists of products of {\sc Nielsen}--integrals up to 
transcendentality four. The computer code {\tt ANCONT} is provided.
\end{abstract}

\end{titlepage}

\newpage
\sloppy
%======================================================================
\begin{minipage}[t]{8cm}{
\vspace*{0.5cm}
\begin{sloppypar}
   {\em Title of program\/}: {\tt ANCONT} \\
   {\em Version\/}: 1.0 \\
   {\em Release\/}: 1  \\
   {\em Catalogue number\/}: \\
   {\em Program obtained from\/}:
   {\tt blumlein@ifh.de} \\
   {\em E-mail\/}: {\tt blumlein@ifh.de} \\
   {\em Licensing provisions\/}: non \\
   {\em Computers\/}: all \\
   {\em Operating system\/}: all \\
   {\em Program language\/}: {\tt FORTRAN77     } \\
   {\em Memory required to execute\/}: Size: 324k, res.~152k \\
   {\em No. of bits per word\/}: up to 32 \\
   {\em No of lines in distributed program\/}: 6696 \\
   {\em Other programs called\/}: non \\
   {\em External files needed\/}: non \\
   {\em Keywords\/}:  Mellin transforms, analytic continuation,
   harmonic sums, polylogarithms, Nielsen functions, Bernoulli numbers, 
   Stirling numbers.
\\
    {\em Nature of the physical problem\/}: \\
Calculation of the analytic continuation of the Mellin transforms to 
complex values of $N$ for the basic functions $g_i(x)$ emerging in 
the splitting and coefficient functions of massless QED and QCD up to 
two--loop order.
\\
   {\em Method of solution\/}: \\
   Serial representations for complex arguments and  approximate 
   polynomial
   representations.\\
   {\em Restrictions on complexity of the problem\/}: non \\
   {\em Typical running time\/}:
From  fractions of a second for individual representations to
3:30 minutes for the test of all options on a HP-9000/755 workstation.\\
\end{sloppypar}}
\end{minipage}
%==============================================================
\newpage

%%%%%%%%%%%%%%%%%%%%%%%%%%%%%%%%%%%%%%%%%%%%%%%%%%%%%%%%%%%%%%%%%%%%%%%
\section{Introduction}
\label{sec:intro}
%%%%%%%%%%%%%%%%%%%%%%%%%%%%%%%%%%%%%%%%%%%%%%%%%%%%%%%%%%%%%%%%%%%%%%%

\vspace{2mm}
\noindent
Hard processes in massless gauge field theories as QED and QCD exhibit
{\sc Mellin}--structures. The observables $\sigma$ can be written as
{\sc Mellin} convolutions
%-----------------------------------------------------------------------
\begin{eqnarray}
\sigma(x) = \left[A \otimes B \right](x) = \int_0^1 dx_1 \int_0^1 dx_2~
\delta(x - x_1 x_2) A_1(x_1) A_2(x_2)~.
\end{eqnarray}
%-----------------------------------------------------------------------
Here, the functions $A_i(x_i)$ denote hard cross sections and splitting
functions, respectively. 
In practice even multiple {\sc Mellin} 
convolutions occur in the calculations of
higher order Feynman diagrams.
Integrals of this type can be diagonalized by the
{\sc Mellin} transform
%-----------------------------------------------------------------------
\begin{eqnarray}
\M\left[\sigma(x)\right](N) = \int_0^1 dx~x^N \sigma(x) =
\M\left[A_1(x_1)\right](N) \cdot
\M\left[A_2(x_2)\right](N)
\end{eqnarray}
%-----------------------------------------------------------------------
and $\sigma(x)$ can be found by a single inverse {\sc Mellin} 
transform after
evaluating  the functions $\M\left[A_i(x_i)\right](N)$, which are simpler
in general. For $N~\epsilon~\N^+$ the {\sc Mellin}
 transforms of the functions
$A_i(x)$ can be represented in terms of
 linear combinations of finite harmonic
sums~[1--3] and their polynomials.
%------------------------------------------------------------------------
\begin{equation}
S_{k_1,\ldots,k_m}(N) = 
\sum_{n_1=1}^N 
\frac{\left[{\rm sign}(k_1)\right]^{n_1}}{n_1^{|k_1|}}
\sum_{n_2=1}^{n_1}
\frac{\left[{\rm sign}(k_2)\right]^{n_2}}{n_2^{|k_2|}} \ldots
\sum_{n_m=1}^{n_{m-1}}
\frac{\left[{\rm sign}(k_m)\right]^{n_m}}{n_m^{|k_m|}}~.
\end{equation}
%------------------------------------------------------------------------
Up to the level of two--loop order
only harmonic sums up to rank 4
contribute. The set of functions $A_i(x)$ is  formed by polynomials
of {\sc Nielsen} integrals~\cite{NIELS}
%------------------------------------------------------------------------
\begin{equation}
\label{eqsnp}
S_{n,p}(x) =  \frac{(-1)^{n+p-1}}{(n-1)!p!} \int_0^1 \frac{dz}{z}
\log^{n-1}(z) \log^p(1-zx)~.
\end{equation}
%------------------------------------------------------------------------
As was shown in Ref.~\cite{BK2} the representation of the {\sc Mellin} 
transforms in terms of finite harmonic sums may be reduced significantly
by applying algebraic relations, cf. also~\cite{BW}, between these sums
which are implied by index permutation and decomposition. Up to the level
of two--loop order
only 25 basic functions remain out of which the
coefficient functions and splitting functions can be assembled as
{\sc Mellin} polynomials.

Any of the multiple {\sc Mellin} convolutions discussed above can thus
be traced back to polynomials of the basic functions and a single
numerical {\sc Mellin} inversion which is performed as a complex
contour integral. The {\sc Mellin}  transforms, which are firstly
obtained at the positive integers, have to be analytically continued
to the complex $N$--plane. This analytic continuation is
unique~\cite{CARLS}. For single harmonic sums the continuation
is well--known
%------------------------------------------------------------------------
\begin{eqnarray}
\label{eqsi1}
S_k(N) &=& (-1)^{k-1} \frac{1}{(k-1)!}
\psi^{(k-1)}(N+1) + c_k^+\\
\label{eqsi2}
S_{-k}(N) &=& (-1)^{k-1+N} \frac{1}{(k-1)!}
\beta^{(k-1)}(N+1) - c_k^-~,
\end{eqnarray}
%------------------------------------------------------------------------
with $\psi(z)$ the logarithmic derivative of the {\sc Euler} 
$\Gamma$--function and
%------------------------------------------------------------------------
\begin{eqnarray}
\label{eqbeta}
\beta(z) &=& \frac{1}{2} \left[ \psi\left(
\frac{z+1}{2}\right)
- \psi\left(\frac{z}{2}\right)\right] \\
 c_1^+ &=& \gamma_E\\
 c_k^+ &=& \zeta(k),~~k \geq 2\\
 c_1^- &=&   \log(2) \\
 c_k^+ &=&  \left(1 -\frac{1}{2^{k-1}}\right)\zeta(k),~~k \geq 2~.
\end{eqnarray}
%------------------------------------------------------------------------
Here $\gamma_E$ denotes the {\sc Euler--Mascheroni} constant and
$\zeta(k)$ is the {\sc Riemann} $\zeta$--function.

It is the aim of the present paper to calculate the analytic continuations
of the {\sc Mellin} transforms of these 25 basic functions and to  
provide a code for numerical evaluations. The paper is organized as
follows. The basic functions are introduced in section~2 and their
structure is discussed. In section~3 the representation of the basic
functions is given for positive integers. These representations which are
given in terms of multiple harmonic sums do also provide first
detailed numerical tests for the representations  used for the
analytic continuations later.
The {\sc Mellin} transforms for complex argument are given in section~4
for the functions of the type $f_i(x)/(1+x)$ and for the
functions $(f_k(x)-f_k(1))/(x-1)$ in section~5. Here we aim on  high
numerical accuracy. Section~6 describes the basic options of the code
{\tt ANCONT}. The results are summarized in
     section~7.
%%%%%%%%%%%%%%%%%%%%%%%%%%%%%%%%%%%%%%%%%%%%%%%%%%%%%%%%%%%%%%%%%%%%%%%
\section{Basic Functions}
\label{sec:basic}
%%%%%%%%%%%%%%%%%%%%%%%%%%%%%%%%%%%%%%%%%%%%%%%%%%%%%%%%%%%%%%%%%%%%%%%

\vspace{2mm}
\noindent
Using the algebraic relations given in Ref.~\cite{BK2} one may show that 
the linear representations up to threefold harmonic sums 
up to 
transcendentality four can be expressed by the single harmonic sums 
$S_{\pm k}(N)$ and the {\sc Mellin}--transforms of the following 25 basic
functions~:
%------------------------------------------------------------------------
\renewcommand{\arraystretch}{2.0}
\begin{equation}
%------------------------------------------------------------------------
\begin{array}{lclcl}
g_1(x) &=& {\ds \frac{\log(1+x)}{x+1}} & \la &S_{1,-1}(N) \\
%------------------------------------------------------------------------
g_2(x) &=& {\ds \frac{\log^2(1+x)}{x+1}} & \la & S_{1,1,-1}(N)\\
%------------------------------------------------------------------------
g_3(x) &=& {\ds \frac{\Li_2(x)}{x+1}}  & \la  & S_{-2,1}(N) \\
%------------------------------------------------------------------------
g_4(x) &=& {\ds \frac{\Li_2(-x)}{x+1}} & \la &  S_{2,-1}(N) \\
%------------------------------------------------------------------------
g_5(x) &=& {\ds \frac{\log(x)\Li_2(x)}{x+1}} &\la & S_{-2,2}(N), 
S_{-3,1}(N)
\\
%------------------------------------------------------------------------
g_6(x) &=& {\ds \frac{\Li_3(x)}{x+1}}  &\la & S_{-3,1}(N)\\
%------------------------------------------------------------------------
g_7(x) &=&{\ds \frac{\Li_3(-x)}{x+1}} & \la &  S_{3,-1}(N) \\
%------------------------------------------------------------------------
g_8(x) &=&{\ds \frac{\Sf(x)}{x+1}}    & \la &  S_{-2,1,1}(N)\\
%------------------------------------------------------------------------
g_9(x) &=&{\ds \frac{\Sf(-x)}{x+1}}   & \la &  S_{2,1,-1}(N)\\
%------------------------------------------------------------------------
g_{10}(x) &=&{\ds \frac{I_1(x)}{1+x}  } & \la &  S_{-1,-2,-1}(N),
S_{2,-1,1}(N)
\\
%------------------------------------------------------------------------
g_{11}(x) &=&{\ds \frac{\log(1-x) \Li_2( x)}{1+x}} & \la &  S_{-1,2,1}(N),
S_{-2,1,1}
\\
%------------------------------------------------------------------------
g_{12}(x) &=&{\ds
\log(1-x) \frac{\Li_2(-x)}{x+1}}  & \la &  S_{-1,-2,-1}(N),
S_{2,-1,1}(N), S_{-2,-1,-1}(N)
\\
%------------------------------------------------------------------------
g_{13}(x) &=&{\ds \frac{\log(1+x)}{1+x} \Li_2( -x)} & \la & 
S_{1,2,-1}(N), S_{2,1,-1}(N)
\\
%------------------------------------------------------------------------
g_{14}(x) &=&{\ds \frac{\log^2(1+x)-\log^2(2)}{x-1}} & \la & 
S_{-1,1,-1}(N)
%------------------------------------------------------------------------
\end{array}
\label{funlist}
\end{equation}
%------------------------------------------------------------------------
\begin{eqnarray}
\begin{array}{lclcl}
%------------------------------------------------------------------------
g_{15}(x) &=&{\ds \frac{\log(1+x) - \log(2)}{x-1} \Li_2(x)} & \la &
S_{-1,-2,1}(N), S_{2,-1,-1}(N), S_{-2,-1,1}(N)
\\
%------------------------------------------------------------------------
g_{16}(x) &=&{\ds \frac{\log(1+x) - \log(2)}{x-1} \Li_2(-x)} & \la &
S_{-1,2,-1}(N), S_{-2,1,-1}(N)
\\
%------------------------------------------------------------------------
g_{17}(x) &=&{\ds \frac{\log(x)\log^2(1+x)}{x-1}} & \la &  
S_{-1,1,-2}(N), S_{-1,2,-1}(N), S_{-2,1,-1}(N)
\\
%------------------------------------------------------------------------
g_{18}(x) &=&{\ds \frac{\Li_2(x)-\zeta(2)}{x-1}} & \la &  S_{2,1}(N) \\
%------------------------------------------------------------------------
g_{19}(x) &=&{\ds \frac{\Li_2(-x)+\zeta(2)/2}{x-1}} & \la & S_{-2,-1}(N)\\
%------------------------------------------------------------------------
g_{20}(x) &=&  {\ds \frac{\Li_3(x)-\zeta(3)}{x-1}}  &  \la & S_{3,1}(N)
\\
%------------------------------------------------------------------------
g_{21}(x) &=&  {\ds \frac{\Sf(x)-\zeta(3)}{x-1}} &  \la &  S_{2,1,1}(N)\\
%------------------------------------------------------------------------
g_{22}(x) &=&{\ds \frac{\log(x)\Li_2(x)}{x-1}} &  \la &  S_{3,1}(N)\\
%------------------------------------------------------------------------
g_{23}(x) &=&{\ds \frac{\Li_3(-x)+3 \zeta(3)/4}{x-1}} &  \la &
S_{-3,-1}(N)\\
%------------------------------------------------------------------------
g_{24}(x) &=&{\ds \frac{I_1(x) + (5/8) \zeta(3)}{x-1}}
& \la &  S_{2,-1,-1}(N), S_{-2,-1,1}(N)
\\
%------------------------------------------------------------------------
g_{25}(x) &=&{\ds \frac{\Sf(-x) - \zeta(3)/8}{x-1}} & \la &
S_{-2,1,-1}(N),  \\
%------------------------------------------------------------------------
\end{array}
\nonumber
\end{eqnarray}
%------------------------------------------------------------------------
\renewcommand{\arraystretch}{1.0}
with
%-----------------------------------------------------------------------
\begin{eqnarray}
\label{eqI1}
I_1(x) = \int_0^x \frac{dz}{z} \log(1-z) \log(1+z)~.
\end{eqnarray}
%-----------------------------------------------------------------------
Here we listed also all harmonic sums of rank~4 which contribute to the
respective {\sc Mellin} transforms.

The functions $g_i(x)$
belong to the class of {\sc Nielsen}--integrals 
\cite{NIELS} ${\rm S}_{n,p}(x)$,~Eq.~(\ref{eqsnp}).
The degree of {\it transcendentality} of these functions is defined to
be $\tau = p+n$. The transcendentality of the product of two of these
these functions is the sum of their transcendentalities. The measures
$dx/(x \pm 1)$ are defined to be of transcendentality 1. The 
polylogarithms are given by
%-----------------------------------------------------------------------
\begin{eqnarray}
\Li_n(x) = \frac{d \Li_{n+1}(x)}{d \log(x)} \equiv  \SN_{n-1,1}(x)
=
\frac{(-1)^{n-1}}{(n-2)!} \int_0^1 \frac{dz}{z}
\log^{n-2}(z) \log(1-zx)~{\rm for}~~{n \geq 2}~.
\end{eqnarray}
%-----------------------------------------------------------------------
The logarithms $\log(1 \pm x)$ are related to the dilogarithm
by~\cite{GRHO}
%-----------------------------------------------------------------------
\begin{eqnarray}
\frac{d \Li_2(\pm x)}{d \log(x)} = \Li_1(\pm x) = - \log(1 \mp x)
\end{eqnarray}
%-----------------------------------------------------------------------
and
%-----------------------------------------------------------------------
\begin{eqnarray}
\Li_0(x) =  \frac{x}{1-x}~.
\end{eqnarray}
%-----------------------------------------------------------------------
Similarly,
%-----------------------------------------------------------------------
\begin{eqnarray}
\frac{d \SN_{n,p}(x)}{d \log(x)} = \SN_{n-1,p}(x)
\end{eqnarray}
%-----------------------------------------------------------------------
holds. As a generalization of the {\sc Nielsen}--integrals~(\ref{eqsnp})
one may wish to consider
%-----------------------------------------------------------------------
\begin{equation}
\label{eqNIE1}
{\rm S}_{n,p,q}(x) =
\frac{(-1)^{n+p+q-1}}{(n-1)! p! q!} \int_0^1 \frac{dz}{z}
\log^{n-1}(z) \log^p(1-zx) \log^q(1+zx)~.
\end{equation}
%-----------------------------------------------------------------------
Both the classes of functions (\ref{eqsnp}) and (\ref{eqNIE1}) are
non--minimal and large subsets of them can be represented already
by the polylogarithms $\Li_l(y)$, where $y$ denotes a
function of $x$. Relations between the different {\sc Nielsen}--integrals
and their generalizations are, however, more easily established
using     {\it non--minimal} representations, since the argument
structure turns out to be simple in the latter case.
Following this line we have introduced in the above set of
functions
%-----------------------------------------------------------------------
\begin{eqnarray}
I_1(x) \equiv S_{1,1,1}(x) =
\int_0^x \frac{dz}{z} \log(1-z) \log(1+z)
= \frac{1}{2} \Sf(x^2) - \Sf(x) - \Sf(-x),
\end{eqnarray}
%-----------------------------------------------------------------------
instead referring to $\Sf(x^2)$.

The {\sc Mellin} transforms of the functions     (\ref{funlist})
are related to a series of finite harmonic sums through which all
harmonic sums up to level three and transcendentality four may be 
expressed in terms of polynomials, except of the well--known case
of single harmonic sums and polynomials which are made of only single
harmonic sums. To establish the connection more closely we mentioned
above
those sums in which the respective {\sc Mellin}  transforms
occur in explicit form, see also Ref.~\cite{BK2}.
%%%%%%%%%%%%%%%%%%%%%%%%%%%%%%%%%%%%%%%%%%%%%%%%%%%%%%%%%%%%%%%%%%%%%%%
\section{Representations of the Mellin Transforms at Positive
Integers}
\label{sec:repi}
%%%%%%%%%%%%%%%%%%%%%%%%%%%%%%%%%%%%%%%%%%%%%%%%%%%%%%%%%%%%%%%%%%%%%%%

\vspace{2mm}
\noindent
To compare the numerical expressions of the analytic continuations of the
{\sc Mellin}--transforms given in the forthcoming sections we summarize 
the exact representations of the {\sc Mellin}--transforms for 
$N~\epsilon~{\bf N^+}$ for the basic functions in terms of harmonic sums.
 These relations are used as one test for the
numerical accuracy of the analytic continuations.
%------------------------------------------------------------------------
\begin{eqnarray}
%------------------------------------------------------------------------
\Mvec\left[g_1(x)\right](N)
&=& (-1)^{N} \Biggl\{
                             S_{-1,1}(N)+\frac{1}{2} \log^2(2)
                             -\left[S_1(N)-S_{-1}(N)\right] \log(2) 
                                     \nonumber\\ & &
                             -S_{1}(N)S_{-1}(N) -S_{-2}(N)\Biggr \}
\\
%------------------------------------------------------------------------
\Mvec\left[g_2(x)\right](N)
&=& 2 (-1)^{N} \Biggl\{S_{1,1,-1}(N) - \log(2)
\left[S_{1,-1}(N) - S_{1,1}(N)\right] \nonumber\\ & &
- \frac{1}{2} \log^2(2) \left[
S_{1}(N)-S_{-1}(N)\right] + \frac{1}{6} \log^3(2)\Biggr \}
\\
%------------------------------------------------------------------------
\Mvec\left[g_3(x)\right](N)
&=& (-1)^{N+1} \left[S_{-2,1}(N)
- \zeta(2) S_{-1}(N) + \frac{5}{8} \zeta(3) - \zeta(2) \log(2) \right]
\\
%------------------------------------------------------------------------
\Mvec\left[g_4(x)\right](N)
&=& (-1)^{N+1} \Biggl\{S_{2,-1}(N) + \log(2)
\left[S_2(N)-S_{-2}(N)\right] + \frac{1}{2} \zeta(2) S_{-1}(N)
\nonumber\\ & &
- \frac{1}{4} \zeta(3) + \frac{1}{2} \zeta(2) \log(2) \Biggr \}
\\
%------------------------------------------------------------------------
\Mvec\left[g_5(x)\right](N)
&=& (-1)^{N} \left[S_{-2,2}(N) + 2 S_{-3,1}(N) - 2 \zeta(2) S_{-2}(N)
- \frac{3}{40} \zeta(2)^2\right]
\\
%------------------------------------------------------------------------
\Mvec\left[g_6(x)\right](N)
&=& (-1)^{N} \Biggl[S_{-3,1}(N)
- \zeta(2) S_{-2}(N) + \zeta(3) S_{-1}(N) \nonumber\\ & &
+ \frac{3}{5} \zeta(2)^2
- 2 \Li_4\left(\frac{1}{2}\right) - \frac{3}{4} \zeta(3) \log(2)
+ \frac{1}{2} \zeta(2) \log^2(2) - \frac{1}{12} \log^4(2) \Biggr ]
\\
%------------------------------------------------------------------------
\Mvec\left[g_7(x)\right](N)
&=& (-1)^{N} \Biggl\{S_{3,-1}(N) + \log(2)
\left[S_3(N) - S_{-3}(N)\right] + \frac{1}{2} \zeta(2) S_{-2}(N)
\nonumber\\ & &
- \frac{3}{4} \zeta(3) S_{-1}(N) + \frac{1}{8} \zeta(2)^2
- \frac{3}{4} \zeta(3) \log(2) \Biggr\}
\\
%------------------------------------------------------------------------
\Mvec\left[g_8(x)\right](N)
&=& (-1)^{N+1} \Biggl
[S_{-2,1,1}(N)
- \zeta(3) S_{-1}(N) + \Li_4\left(\frac{1}{2}\right) 
- \frac{1}{8} \zeta(2)^2 - \frac{1}{8} \zeta(3) \log(2) \nonumber \\ & &
- \frac{1}{4}
\zeta(2) \log^2(2) + \frac{1}{24} \log^4(2)\Biggr ]
\\
%------------------------------------------------------------------------
\Mvec\left[g_9(x)\right](N) &=&
(-1)^{N+1} 
\Biggl\{S_{2,1,-1}(N)+\log(2)\left[S_{2,1}(N)-S_{2,-1}(N)\right]
\nonumber\\ & &
- \frac{1}{2}  \log^2(2) \left[S_2(N)-S_{-2}(N)\right]
- \frac{1}{8} \zeta(3) S_{-1}(N)
\nonumber\\ & &
- 3 \Li_4\left(\frac{1}{2}\right) + \frac{6}{5} \zeta(2)^2 - \frac{11}{4}
\zeta(3) \log(2) + \frac{3}{4} \zeta(2) \log^2(2) - \frac{1}{8} \log^4(2)
\Biggr\}
\\
%------------------------------------------------------------------------
\Mvec\left[g_{10}(x)\right](N) &=& (-1)^{N+1} \Biggl\{S_{-2,-1,-1}(N)
+S_{2,-1,1}(N) - \log(2) \left[S_{-2,1}(N)-S_{-2,-1}(N)\right]
\nonumber\\ & &
+ \frac{1}{2}\left[\zeta(2) - \log^2(2)\right] \left[S_2(N)-S_{-2}(N)
\right] + \frac{5}{8} \zeta(3) \left[S_{-1}(N)+\log(2)\right] \nonumber\\
& &
-
\frac{3}{20} \zeta(2)^2  \Biggr\}
\\
%------------------------------------------------------------------------
\Mvec\left[g_{11}(x)\right](N) &=&
(-1)^N \Biggl\{S_{-1,2,1}(N) + 2 S_{-2,1,1}(N) - 2 \zeta(3) S_{-1}(N)
- \zeta(2) S_{-1,1}(N) \nonumber\\ & &
- \frac{29}{40} \zeta(2)^2 + 3 \Li_4\left(\frac{1}{2}\right)
- \frac{1}{4} \zeta(2) \log^2(2) + \frac{1}{4} \log^4(2) \Biggr\}
\\
%------------------------------------------------------------------------
\Mvec\left[g_{12}(x)\right](N) &=&  (-1)^N  \Biggl\{
S_{2,-1,1}(N)+S_{-2,-1,-1}(N)+S_{-1,-2,-1}(N) \nonumber\\ & &
-\log(2)
\left[S_{-2,1}(N)+ S_{-1,2}(N)\right]
+\frac{1}{2} \zeta(2) S_{-1,1}(N)\nonumber\\ & &
+\left[S_{-1}(N) S_{-2}(N)+S_3(N)
\right]\log(2)\nonumber\\ & &
+\frac{1}{2} \left[\zeta(2)-\log^2(2)\right] \left[S_2(N)-S_{-2}(N)\right
]+\frac{5}{8} \zeta(3) S_{-1}(N) \nonumber\\ & &
-4 \Li_4\left(\frac{1}{2}\right) + \frac{3}{2} \zeta)2)^2 - \frac{21}
{8} \zeta(3) \log(2)  + \frac{3}{4} \zeta(2) \log^2(2)
-\frac{1}{6} \log^4(2) \Biggr\}
\\
%------------------------------------------------------------------------
\Mvec\left[g_{13}(x)\right](N) &=&
(-1)^N\Biggl\{S_{1,2,-1}(N)+2S_{2,1,-1}(N)+\frac{1}{2} \zeta(2)
S_{1,-1}(N) \nonumber\\ & &
- \log^2(2) \left[S_2(N)-S_{-2}(N)\right] \nonumber\\ & &
+ \log(2) \Bigl[S_{2,1}(N)-S_{1,-2}(N)-2 S_{2,-1}(N)+S_1(N) S_2(N)
\nonumber\\ & &
+S_3(N)- \frac{1}{2} \zeta(2) S_{-1}(N) \Bigr ] \nonumber\\ & &
- 3 \Li_4\left(\frac{1}{2}\right)+\frac{6}{5} \zeta(2)^2 -
\frac{21}{8} \zeta(3) \log(2) +\frac{1}{2} \zeta(2) \log^2(2)
- \frac{1}{8} \log^4(2) \Biggr\}
\\
%------------------------------------------------------------------------
\Mvec\left[g_{14}(x)\right](N) &=&
2 S_{-1,1,-1}(N) - 2 \log(2) \left[S_{-1,-1}(N)-S_{-1,1}(N)\right]
- \log^2(2) S_{-1}(N) \nonumber\\ & &
- \frac{1}{4} \zeta(3) + \zeta(2) \log(2) - \frac{2}{3} \log^3(2)
\\
%------------------------------------------------------------------------
\Mvec\left[g_{15}(x)\right](N) &=& S_{-1,-2,1}(N)+S_{2,-1,-1}(N)
+S_{-2,-1,1}(N)-\zeta(2) S_{-1,-1}(N) + \frac{5}{18} \zeta(3) S_{-1}(N)
\nonumber\\ & &
- \left[\frac{5}{8} \zeta(3)
-\zeta(2) \log(2)\right]\left[S_1(N)-S_{-1}(N)\right]
+ \log(2) \left[S_{2,1}(N) -2 \zeta(3)\right] 
\nonumber\\ &&
-\frac{1}{2} \left(\zeta(2) - \log^2(2)\right) \left[S_2(N)-S_{-2}(N)
\right] - \log(2) \zeta(2) S_1(N) 
\nonumber\\ & &
+ \frac{19}{4} \zeta(2)^2 - \Li_4\left(\frac{1}{2}\right)
+ \frac{7}{4} \zeta(3) \log(2) - \frac{1}{4} \zeta(2) \log^2(2)
-\frac{1}{24} \log^4(2)
\\
%------------------------------------------------------------------------
\Mvec\left[g_{16}(x)\right](N) &=&
2 S_{-2,1,-1}(N) + S_{-1,2,-1}(N) \nonumber\\ & &
+ \log(2)\left[2 \left(S_{-2,1}(N)-S_{-2,-1}(N)\right)+S_{-1,2}(N)
-S_{-1,-2}(N)\right]\nonumber\\ & &
+ \frac{1}{2} \zeta(2) S_{-1,-1}(N) - \log^2(2) \left[S_{-2}(N)
-S_{2}(N)\right] - \frac{1}{4}  \zeta(3) S_1(N) \nonumber\\ & &
-\left(\frac{1}{4} \zeta(3) - \frac{1}{2} \zeta(2) \log(2) \right)
\left[S_{-1}(N) -S_1(N)\right] \nonumber\\ & &
+ \log(2) \left[S_{-2,-1}(N) - \log(2) \left(S_2(N)-S_{-2}(N)\right)
+ \frac{1}{2} \zeta(2) S_1(N) \right] \nonumber\\ & &
- \frac{33}{20} \zeta(2)^2 + 4 \Li_4\left(\frac{1}{2}\right)
+ \frac{13}{4} \zeta(3) \log(2) - \frac{3}{4} \zeta(2) \log^2(2)
+ \frac{1}{6} \log^4(2)
\\
%------------------------------------------------------------------------
\Mvec\left[g_{17}(x)\right](N) &=& - 2 \left[
 S_{-1,2,-1}(N)+S_{-1,1,-2}(N)+S_{-2,1,-1}(N)\right]
\nonumber\\ & &
+   \log(2) \left[S_{-1,-2}(N)-S_{-1,2}(N) - S_{-2,1}(N) +S_{-2,-1}(N)
\right]
\nonumber\\ & &
+ \frac{1}{8} \zeta(3)       S_{-1}(N)
+ \frac{1}{2}
\log^2(2)            \left[S_{-2}(N)-S_2(N)\right] - \frac{1}{4} \zeta(2)
S_{-1,1}(N)
\nonumber\\ & &
+ \frac{7}{4} \zeta(2)^2 - 4 \Li_4\left(\frac{1}{2}\right)
- \frac{21}{4} \zeta(3) \log(2) + \frac{5}{2} \zeta(2) \log^2(2)
-\frac{1}{6} \log^4(2)
\\
%------------------------------------------------------------------------
\Mvec\left[g_{18}(x)\right](N) &=&
-S_{2,1}(N)+2\zeta(3)
\\
%------------------------------------------------------------------------
\Mvec\left[g_{19}(x)\right](N) &=&
-S_{-2,-1}(N)+\log(2)\left[S_2(N)-S_{-2}(N)\right] - \frac{5}{8} \zeta(3)
\\
%------------------------------------------------------------------------
\Mvec\left[g_{20}(x)\right](N) &=&
  S_{3,1}(N) - \zeta(2) S_2(N)+\frac{1}{2} \zeta(2)^2
\\
%------------------------------------------------------------------------
\Mvec\left[g_{21}(x)\right](N) &=& -S_{2,1,1}(N)+\frac{6}{5} \zeta(2)^2
\\
%------------------------------------------------------------------------
\Mvec\left[g_{22}(x)\right](N) &=& 
   2 S_{3,1}(N)+ \frac{1}{2} S_{4}(N) + \frac{1}{2} S_2^2(N)
- 2 \zeta(2) S_2(N) + \frac{3}{10} \zeta(2)^2
\\
%------------------------------------------------------------------------
\Mvec\left[g_{23}(x)\right](N) &=&
 S_{-3,-1}(N)-\log(2) \left[S_3(N)-S_{-3}(N)\right]
+\frac{1}{2} \zeta(2) S_2(N) \nonumber\\ & &
+2 \Li_4\left(\frac{1}{2}\right)  - \frac{11}{10} \zeta(2)^2 + \frac{7}
{4} \zeta(3) \log(2) - \frac{1}{2} \zeta(2) \log^2(2)
+ \frac{1}{12} \log^4(2)
\\
%------------------------------------------------------------------------
\Mvec\left[g_{24}(x)\right](N) &=&
-S_{2,-1,-1}(N)-S_{-2,-1,1}(N)+\log(2)\left[S_{2,1}(N)-
S_{2,-1}(N)\right]\nonumber\\ & &
+\frac{1}{2}\left[\zeta(2)-\log^2(2)\right]
\left[S_2(N)-S_{-2}(N)\right] - \frac{5}{8} \zeta(3) S_1(N)
\nonumber\\ & & + \frac{1}{4} \zeta(2)^2-2 \Li_4\left(\frac{1}{2}\right)
- \frac{7}{4} \zeta(3) \log(2) + \frac{1}{2} \zeta(2) \log^2(2)
- \frac{1}{12} \log^4(2)
\\
%------------------------------------------------------------------------
\Mvec\left[g_{25}(x)\right](N) &=&
- S_{-2,1,-1}(N) - \log(2) \left[S_{-2,1}(N) - S_{-2,-1}(N)\right]
\nonumber\\ & &
+ \frac{1}{2} \log^2(2) \left[S_{-2}(N)-S_2(N)\right] + \frac{3}{40}
\zeta(2)^2
\end{eqnarray}
%------------------------------------------------------------------------
The above relations can be checked by numerical integration directly.
In the code {\tt ANCONT} we use the routines {\tt DAIND}, 
Ref.~\cite{AIND}, and verified these relations numerically for the
integer moments $N~\epsilon~[1,20]$ at an accuracy of better than
at least  $4 \cdot 10^{-9}$, and in many cases even of
$O(10^{-12})$.
%%%%%%%%%%%%%%%%%%%%%%%%%%%%%%%%%%%%%%%%%%%%%%%%%%%%%%%%%%%%%%%%%%%%%%%
\section{Mellin Transforms for the
Functions $\mathbf{f(x)  /(x+1)}$}
\label{sec:mellinp}
%%%%%%%%%%%%%%%%%%%%%%%%%%%%%%%%%%%%%%%%%%%%%%%%%%%%%%%%%%%%%%%%%%%%%%%

\vspace{2mm}
\noindent
For functions of the type
%----------------------------------------------------------------------
\begin{equation}
g_i(x) = \frac{f_i(x)}{x+1}
\end{equation}
%----------------------------------------------------------------------
and $f_i(x)~\epsilon~{\cal C}^{\infty}[0,1[$,~cf.~\cite{YOS}, one may
expand $f_i(x)$ into a Taylor series around $x=0$,
%----------------------------------------------------------------------
\begin{equation}
f_i(x) = \sum_{k=0}^{\infty} a_k x^k~.
\end{equation}
%----------------------------------------------------------------------
The {\sc Mellin}--transform is then given by
%----------------------------------------------------------------------
\begin{equation}
\Mvec \left[\frac{f_i(x)}{x+1}\right](N) = \sum_{k=0}^{\infty} a_k
\beta(N+k+1)~.
\end{equation}
%----------------------------------------------------------------------
 Since the {\sc Mellin}--transforms of the type
$\Mvec[\log^m(x) f(x)](N)$ can be calculated using the relation
%-----------------------------------------------------------------------
\begin{eqnarray}
\label{deriv}
\Mvec\left[\log^m(x) f(x) \right](N) = \frac{\partial^N}{\partial N^k}
\Mvec\left[f(x)\right](N)~,
\end{eqnarray}
%-----------------------------------------------------------------------
one obtains
%----------------------------------------------------------------------
\begin{equation}
\Mvec\left[\log^m(x) \frac{f_i(x)}{1+x}\right](N)
= \sum_{k=0}^{\infty} a_k \beta^{(m)}(N+k+1)~.
\end{equation}
%----------------------------------------------------------------------
The function $\beta^{(m)}(z)$ obeys the recursion relation
%----------------------------------------------------------------------
\begin{equation}
\beta^{(m)}(z+1) = (-1)^m m! \left[\frac{1}{(1+z)^{m+1}} 
- \frac{1}{z^{m+1}}\right]
+ \beta^{(m)}(z)~.
\end{equation}
%----------------------------------------------------------------------
For the polylogarithms
$\Li_l(\pm x)$ one obtains
%----------------------------------------------------------------------
\begin{equation}
\Mvec \left[\log^m(x) \frac{\Li_l(\pm x)}{x+1}\right](N)
= \sum_{k=1}^{\infty} \frac{(\pm 1)^k}{k^l} \beta^{(m)}(N+k+1)~,
\end{equation}
%----------------------------------------------------------------------
as an example. Although these representations  are quite general
and may be applied to other {\sc Nielsen}--integrals as well,
the corresponding series may not converge fast enough.

Alternatively,
the analytic continuation of the {\sc Mellin}  transforms containing
the factor \mbox{ $(x+1)^{-1}$}
 can be performed using the transformation
%-----------------------------------------------------------------------
\begin{eqnarray}
\label{mxp1}
\Mvec\left[\frac{f(x)}{1+x}\right](N) \equiv
\int_0^1 dx \frac{x^{N}  }{x+1} f(x) &=&
\log(2) \cdot f(1) - \int_0^1 dx x^{N-1}~\log(1+x) \left [ N f(x)
 + x f'(x)\right] \nonumber\\
&=& \log(2) \cdot f(1) - \sum_{k=1}^9 a_k^{(1)} \left\{ N
\Mvec[f(x)](N+k-1) \right. \nonumber 
\\ & & \left. ~~~~~~~~~~~~~~~~~~~~~~~~~~~~~
+ \Mvec[f'(x)](N+k)\right\}~,
\end{eqnarray}
%-----------------------------------------------------------------------
if $f(x)$ is differentiable in $]0,1[$. The function $\log(1+x)$ can
be approximated by
%----------------------------------------------------------------------
\begin{equation}
\label{leq1}
\ln(1+x) \simeq  \sum_{k=1}^{9}a_{k}^{(1)}x^{k}~,
\end{equation}
%-----------------------------------------------------------------------
with an accuracy better than
$3 \times 10^{-8}$. The polynomial is determined using the
approximation given by the
{\tt MINIMAX}--method~\cite{MAPLE}
with an
adaptive  choice of arguments $x$.
The coefficients $a_k^{(1) }$ are given by
%-----------------------------------------------------------------------
\small
\begin{eqnarray}
%NEW
\begin{array}{lcrlcr}
a_1^{(1)}   &=&   0.999999974532238\E+0~~~&
a_2^{(1)}   &=&  -0.499995525889840\E+0
\nonumber\\
a_3^{(1)}   &=&   0.333203435557262\E+0~~~ &
a_4^{(1)}   &=& -0.248529457782640\E+0
\nonumber\\
a_5^{(1)}   &=&   0.191451164719161\E+0~~~&
a_6^{(1)}    &=& -0.137466222728331\E+0
\nonumber\\
a_7^{(1)}  &=&    0.792107412244877\E-1~~~&
a_8^{(1)}  &=&   -  0.301109656912626\E-1
\nonumber\\
a_9^{(1)}  &=&   0.538406208663153\E-2~~~ & & &
\end{array}
\nonumber
\end{eqnarray}
\normalsize
%-----------------------------------------------------------------------
Similar representations are found in the literature
cf.~[11--13]
%cf.~\cite{HAST,ABST,LCY}
and have been used in Ref.~\cite{BK1} before.
Up to two--loop order also the representations of $\log^k(1+x)$ are
needed with $k = 2,3$,
%----------------------------------------------------------------------
\begin{eqnarray}
\label{leq1a}
\ln^2(1+x) &\simeq&  \sum_{k=2}^{11}a_{k}^{(2)}x^{k}~,              
\\
\label{leq1b}
\ln^3(1+x) &\simeq&  \sum_{k=3}^{13}a_{k}^{(3)}x^{k}~.
\end{eqnarray}
%-----------------------------------------------------------------------
The coefficients are
\small
%-----------------------------------------------------------------------
\begin{eqnarray}
%NEW
\begin{array}{lcrlcr}
a_2^{(2)}   &=&0.999999980543793\E+0~~~&
a_3^{(2)}   &=&-0.999995797779624\E+0
\nonumber\\
a_4^{(2)}   &=&0.916516447393493\E+0~~~&
a_5^{(2)}   &=&-0.831229921350708\E+0
\nonumber\\
a_6^{(2)}   &=& 0.745873737923571\E+0~~~&
a_7^{(2)}    &=&-0.634523908078600\E+0
\nonumber\\
a_8^{(2)}  &=&    0.467104011423750\E+0~~~&
a_9^{(2)}  &=&   -0.261348046799178\E+0
\nonumber\\
a_{10}^{(2)}  &=&  0.936814286867420\E-1~~~&
a_{11}^{(2)}  &=&  -0.156249375012462\E-1
\end{array}
\nonumber
\end{eqnarray}
%-----------------------------------------------------------------------
\normalsize
and
\small
%-----------------------------------------------------------------------
\begin{eqnarray}
%NEW
\begin{array}{lcrlcr}
a_3^{(3)}   &=&0.999999989322696\E+0~~~&
a_4^{(3)}   &=&-0.149999722020708\E+1
\nonumber\\
a_5^{(3)}   &=&0.174988008499745\E+1~~~ &
a_6^{(3)}   &=&-0.187296689068405\E+1
\nonumber\\
a_7^{(3)}   &=&0.191539974617231\E+1~~~&
a_8^{(3)}    &=&-0.185963744001295\E+1
\nonumber\\
a_9^{(3)}  &=& 0.162987195424434\E+1~~~&
a_{10}^{(3 )}  &=&-0.117982353224299\E+1
\nonumber\\
a_{11}^{(3)}  &=&0.628710122994999\E+0~~~&
a_{12}^{(3)}  &=&-0.211307487211713\E+0
\nonumber\\
a_{13}^{(3)}  &=& 0.328953352932140\E-1~~~
&  & &
\end{array}
\nonumber
\end{eqnarray}
%-----------------------------------------------------------------------
\normalsize
Eqs.~(\ref{leq1a},\ref{leq1b}) hold
at an accuracy better than $2 \times 10^{-8}$ for $x~\epsilon~[0,1]$.

The {\sc Mellin}--transforms of the functions
$\log(1+x)/(1+x)$ and
$\log^2(1+x)/(1+x)$ are obtained by
%-----------------------------------------------------------------------
\begin{eqnarray}
\label{eqlogk}
\Mvec\left[\frac{\log^{k}(1+x)}{1+x}\right](N) &=&
 \frac{1}{k+1} \left[
\log^{k+1}(2) - N
 \Mvec\left[\log^{k+1}(1+x)\right](N-1)\right]
\nonumber\\
&=&  \frac{1}{k+1} \left[
\log^{k+1}(2) - N
  \sum_{l=k+1}^{L_1(k+1)} \frac{a_l^{(k+1)}}{N+l}\right]~.
\end{eqnarray}
%-----------------------------------------------------------------------
with $L_1(2) = 11$ and $L_1(3) = 13$.
Similarly
the {\sc Mellin}--transform of the function $g_{17}(x)$
is obtained by
%-----------------------------------------------------------------------
\begin{equation}
\Mvec\left[\frac{\log(x)}{x-1} \log^2(1+x)\right](N) =
           \sum_{k=2}^{11} a_k^{(2)} \psi'(N+k+1)~.
\end{equation}
%-----------------------------------------------------------------------
The remaining {\sc Mellin}--transforms are obtained from the 
representation
   Eq.~(\ref{mxp1}) and are expressed in terms of the known
{\sc Mellin}--transforms for the functions $f(x)$ and 
$f'(x)$,~cf.~\cite{BK2}~:
%-----------------------------------------------------------------------
\begin{eqnarray}
\Mvec\left[\Li_2(x)\right](N) &=& \frac{1}{N+1}\left[\zeta(2) - \frac{
S_1(N+1)}{N+1}\right] \\
%----
\Mvec\left[\Li'_2(x)  \right](N) &=& - \Mvec\left[\log(1-x)\right](N-1)
= \frac{S_1(N)}{N}    
\end{eqnarray}
%-----------------------------------------------------------------------
etc.
The single harmonic sums
$S_{\pm k}(N)$ are given in Eqs.~(\ref{eqsi1},\ref{eqsi2}).\\
One obtains the representations~:
%-----------------------------------------------------------------------
\begin{eqnarray}
\label{mxp2}
\Mvec\left[\frac{\Li_2(x)}{1+x}\right](N) &=& \zeta(2) \log(2)
- \sum_{k=1}^9 a_k^{(1)} 
\left[\frac{N}{N+k} \zeta(2) + \frac{k}{(N+k)^2}
S_1(N+k)\right]
\\
%-----------------------------------------------------------------------
\Mvec\left[\frac{\Li_2(-x)}{1+x}\right](N) 
&=&  - \frac{1}{2} \zeta(2) \log(2) + \sum_{k=1}^9 a_k^{(1)}
\Biggl\{ 
\frac{N}{N+k} \frac{\zeta(2)}{2} + \frac{k}{(N+k)^2} \left[
\log(2) \right. \nonumber\\ & & \left.
- \beta(N+k+1)\right]\Biggr\} \\
%-----------------------------------------------------------------------
\Mvec\left[\log(x) \frac{\Li_2(x)}{1+x}\right](N)
&=& - \sum_{k=1}^9 \frac{k}{(N+k)^2}\left[\zeta(2) + \psi'(k+1) - 2
\frac{S_1(N+k)}{N+k} \right] \\
%-----------------------------------------------------------------------
\Mvec\left[\frac{\Li_3(x)}{1+x}\right](N)
&=& \zeta(3) \log(2) \nonumber\\ & &
- \sum_{k=1}^9 a_k^{(1)} \left\{\frac{N}{N+k}
\zeta(3) + \frac{k}{(N+k)^2} \left[\zeta(2) - \frac{S_1(N+k)}{N+k}\right]
\right\}\\
%-----------------------------------------------------------------------
\Mvec\left[\frac{\Li_3(-x)}{1+x}\right](N)
    &=& - \frac{3}{4} \zeta(3) \log(2) + \sum_{k=1}^9 a_k^{(1)} \left\{
\frac{N}{N+k} \frac{3}{4} \zeta(3) + \frac{k}{(N+k)^2} \frac{1}{2} 
\zeta(2) \right. \nonumber\\ & & \left.
- \frac{k}{(N+k)^3}\left[\log(2) - \beta(N+k+1)\right]\right\}\\
%-----------------------------------------------------------------------
\Mvec\left[\frac{\Sf(x)}{1+x}\right](N)
&=& \log(2) \zeta(3)
- \sum_{k=1}^9 a_k^{(1)} \left \{\frac{N}{N+k} \zeta(3) \right. 
\nonumber\\ & & \left.
+
\frac{k}{(N+k)^2} \frac{1}{2} \left[S_1^2(N+k) + S_2(N+k)\right]\right\}
\\
%-----------------------------------------------------------------------
\Mvec\left[\frac{\Sf(-x)}{1+x}\right](N)
 &=& \frac{1}{8} \zeta(3) \log(2) - \sum_{k=1}^9 a_k^{(1)} 
\frac{N}{N+k} \left[ \frac{\zeta(3)}{8} - \frac{1}{2} \sum_{l=2}^{11}
\frac{a_l^{(2)}}{N+k+l}\right] \nonumber\\ & &
- \frac{1}{2} \sum_{k=3}^{13} 
\frac{a_k^{(3)}}{N+k}
\end{eqnarray}
%-----------------------------------------------------------------------

The function $\Sf    (x^2)$ does not occur singly but rather in the
combination $\Sf(x^2)/2 - \Sf(x) - \Sf(-x)$ in the 
{\sc Mellin}--transforms which are related to the harmonic sums up
to the level being investigated in the present paper. It         emerges
via the integral
%------------------------------------------------------------------------
\begin{equation}
I_1(x)=
\int_0^x \frac{dz}{z} \log(1-z) \log(1+z) = \frac{1}{2} \Sf(x^2)
- \Sf(x) - \Sf(-x)~.
\end{equation}
%------------------------------------------------------------------------
We firstly   calculate the {\sc Mellin}--transform for this integral.
The {\sc Mellin}--transform for $\Sf(x^2)$ is then easily obtained by a
linear combination using the relations given above.
%-----------------------------------------------------------------------
\begin{eqnarray}
\Mvec\left[\frac{I_1(x)}{1+x}\right](N)
&=& - \frac{5}{8} \zeta(3) \log(2) + \sum_{k=2}^{11} a_k^{(2)} 
\frac{S_1(N+k)}{N+k}
\nonumber\\ & &
+ \sum_{k=1}^9 a_k^{(1)} \frac{N}{N+k} \left[
\frac{5}{8} \zeta(3) - \sum_{l=1}^9 a_l^{(1)} \frac{S_1(N+k+l)}{N+k+l}
\right]~.
\end{eqnarray}
%------------------------------------------------------------------------
%%%%%%%%%%%%%%%%%%%%%%%%%%%%%%%%%%%%%%%%%%%%%%%%%%%%%%%%%%%%%%%%%%%%%%%%%
\renewcommand{\arraystretch}{1.3}
\begin{center}
\footnotesize
\begin{tabular}{||r||  r|r|r||}
\hline \hline
\multicolumn{1}{||c}{$N$}&
\multicolumn{1}{|c}{Eq.~(\ref{eqgrv1})}&
\multicolumn{1}{|c}{Eq.~(\ref{eqgrv2})}&
\multicolumn{1}{|c||}{Eq.~(\ref{mxp2})}\\
\hline \hline
0 &
-0.717E-03 &
 0.969E-04 &
 0.212E-08 \\
1 &  
-0.117E-02 &
 0.205E-03 &
 0.442E-08   \\
2 &  
-0.160E-02 &
 0.292E-03 &
 0.761E-08 \\
3 &
-0.207E-02 &
 0.378E-03   &
 0.117E-07 \\
4 &
-0.257E-02 &
 0.452E-03 &
 0.169E-07 \\
5 &
-0.308E-02 &
 0.505E-03 &
 0.232E-07\\
6 &
-0.359E-02 &
 0.531E-03 &
 0.307E-07 \\
7 &
-0.410E-02 &
 0.529E-03 &
 0.397E-07 \\
8 &
-0.461E-02 &
 0.501E-03 &
 0.501E-07 \\
9 &
-0.511E-02 &
 0.449E-03 &
 0.622E-07 \\
10 &
-0.559E-02 &
 0.376E-03 &
 0.758E-07 \\
11 &
-0.607E-02 &
 0.286E-03 &
 0.909E-07 \\
12 &
-0.652E-02 &
 0.181E-03 &
 0.107E-06 \\
13 &
-0.697E-02 &
 0.637E-04 &
 0.125E-06 \\
14 &
-0.740E-02 &
-0.630E-04 &
 0.144E-06 \\
15 &
-0.782E-02 &
-0.197E-03 &
 0.164E-06 \\
16 &
-0.823E-02 &
-0.338E-03 &
 0.184E-06 \\
17 &
-0.862E-02 &
-0.483E-03 &
 0.205E-06 \\
18 &
-0.900E-02 &
-0.632E-03 &
 0.226E-06 \\
19 &
-0.936E-02 &
-0.784E-03 &
 0.247E-06 \\
20 &
-0.972E-02 &
-0.936E-03 &
 0.268E-06 \\
\hline \hline
\end{tabular}
\normalsize
\end{center}

\vspace{2mm}
\noindent
{\sf Table~1: Comparison of the relative accuracy of different
approximations for $\Mvec\left[\Li_2(x)/(1+x)\right](N)$.}
%%%%%%%%%%%%%%%%%%%%%%%%%%%%%%%%%%%%%%%%%%%%%%%%%%%%%%%%%%%%%%%%%%%%%%%
\renewcommand{\arraystretch}{1}

\vspace{4mm}

Since the harmonic sum $S_{-2,1}(N)$ contributes already to the
next--to--leading order anomalous dimensions, approximations for
$\Mvec\left[\Li_2(x)/(1+x)\right
](N)$ have been used in the literature for
some time~\cite{GRV,BK1}.
Two approximations  used before are
%-----------------------------------------------------------------------
\begin{eqnarray}
\label{eqgrv1}
\Mvec\left[\frac{\Li_2(x)}{1+x}\right](N) &\simeq& 
\frac{1.01}{N+2}-\frac{0.846}{N+3}+\frac{1.155}{N+4}-\frac{1.074}{N+5}
+\frac{0.55}{N+6}
\end{eqnarray}
%-----------------------------------------------------------------------
or
%-----------------------------------------------------------------------
\begin{eqnarray}
\label{eqgrv2}
\Mvec\left[\frac{\Li_2(x)}{1+x}\right](N) &\simeq& 
\frac{1.004}{N+2}-\frac{0.846}{N+3}+\frac{1.342}{N+4}-\frac{1.532}{N+5}
+\frac{0.839}{N+6}~,
\end{eqnarray}
%-----------------------------------------------------------------------
cf.~\cite{GRV},
which can be compared to our representation Eq.~(\ref{mxp2}).
In Table~1 the integer moments of these representations are given for
$N~\epsilon~[0,20]$. Eq.~(\ref{mxp2}) is more precise than 
Eqs.~(\ref{eqgrv1},\ref{eqgrv2}) by three to four orders
of magnitude.       
Other quantities were approximated by similar
representations as (\ref{eqgrv1},\ref{eqgrv2}) in~\cite{NV} recently.

A {\sc Mellin}--transform of a function carrying the factor $\log(1-x)/
(1+x)$ can be performed if the remainder factor obeys a polynomial 
representation, since the former factor may be transformed into
the {\sc Mellin}--transform of $\log(1+x)/(1+x)$, 
cf. Ref.~\cite{BK2}. Eqs.~(44,45),
by
%-----------------------------------------------------------------------
\begin{equation}
\label{eqlotra1}
\Mvec\left[\frac{\log(1-x)}{1+x}\right](N) =
 - \Mvec\left[\frac{\log(1+x)}{1+x}\right](N) - \beta(N+1) \left[
\psi(N+1) + \gamma_E - \log(2) \right] + \beta'(N+1).
\end{equation}
%-----------------------------------------------------------------------
Using Eq.~(\ref{eqlogk}) one obtains
%-----------------------------------------------------------------------
\begin{eqnarray}
\Mvec\left[\frac{\log(1-x) \Li_2(-x)}{1+x}
\right](N)
&=&  \frac{1}{2} \sum_{k=1}^9 \frac{a_k^{(1)}}{k}\Biggl\{
\left[
\log^2(2) -
  \sum_{l=2}^{11} a_l^{(2)} \frac{N+k}{N+k+l}\right]
- \beta'(N+k+1)       .
        \nonumber\\ & &
+ \beta(N+k+1)\left[S_1(N+k)-\log(2) \right]\Biggr\}~.
\end{eqnarray}
%-----------------------------------------------------------------------

The {\sc Mellin}--transform of the function $\log(1-x) \Li_2(x)/(1+x)$
can be calculated similarly to the case which was discussed before.
                             $\Li_2(x)$ may be represented by
%-----------------------------------------------------------------------
\begin{equation}
\label{eqLI2a}
\Li_2(x) \simeq  P_0^{(1)}(x)+ \left[P_1^{(1)}(x)
+
\log(1-x) P_2^{(1)}(x)\right],
\end{equation}
%-----------------------------------------------------------------------
where $P_{0,1,2}^{(1)}(x)$ are the
polynomials  
%-----------------------------------------------------------------------
%-----------------------------------------------------------------------
\begin{eqnarray}
\label{eqP0}
P_0^{(1)}(x) &=& \sum_{k=0}^{11} c_k^{(1)} x^k\\
P_1^{(1)}(x) &=& 
- \frac{49}{36} + \zeta(2) + \frac{11}{6} x - \frac{7}{12}
x^2 +\frac{1}{9} x^3 \\
P_2^{(1)}(x) &=& 
\frac{11}{6} - 3 x  + \frac{3}{2} x^2 - \frac{1}{3} x^3~.
\end{eqnarray}
%-----------------------------------------------------------------------
with
%-----------------------------------------------------------------------
\small
\begin{eqnarray}
%NEW
\begin{array}{lcrlcr}
c_0^{(1)} &=&   -0.283822933724932\E+0~~~&
c_1^{(1)} &=&    0.999994319023731\E+0
\nonumber\\
c_2^{(1)} &=&   -0.124975762907682\E+1~~~&
c_3^{(1)} &=&    0.607076808008983\E+0
\nonumber\\
c_4^{(1)} &=&   -0.280403220046588\E-1~~~&
c_5^{(1)} &=&   -0.181869786537805\E+0
\nonumber\\
c_6^{(1)} &=&    0.532318519269331\E+0~~~&
c_7^{(1)} &=&   -0.107281686995035\E+1
\nonumber\\
c_8^{(1)} &=&    0.138194913357518\E+1~~~&
c_9^{(1)} &=&   -0.111100841298484\E+1
\nonumber\\
c_{10}^{(1)} &=& 0.506649587198046\E+0~~~&
c_{11}^{(1)} &=&-0.100672390783659\E+0
\nonumber\\
\end{array}
\nonumber
\end{eqnarray}
\normalsize
%-----------------------------------------------------------------------
The representation Eq.~(\ref{eqLI2a}) holds at an accuracy of better
than $3  \times 10^{-8}$ for $x~\epsilon~[0,1]$.

The {\sc Mellin} transform of $\log(1-x) \Li_2(x)/(1+x)$ is thus
expressed as a sum over {\sc Mellin} transforms 
$\Mvec[\log(1-x)/(1+x)]N)$, Eq.~(\ref{eqlotra1}) and
$\Mvec[\log^2(1-x)/(1+x)](N)$. The latter expression
is related to the harmonic sum $S_{-1,1,1}(N)$,~\cite{BK2},~Eq.~(60).
This sum can be expressed by the sum $S_{1,1,-1}(N)$ and lower order
harmonic sums through
%-----------------------------------------------------------------------
\begin{eqnarray}
\label{eqperm1}
S_{-1,1,1}(N) &=& S_{1,1,-1}(N) + S_1(N) S_{1,-1}(N) + S_{-2,1}(N) +
S_{-1,2}(N) \nonumber\\ & &
+ \frac{1}{2} \left[S_1^2(N) S_{-1}(N) - S_{-1}(N) S_2(N)
\right] - S_{-3}(N).
\end{eqnarray}
%-----------------------------------------------------------------------
The harmonic sum $S_{1,1,-1}(N)$ is related to
$\Mvec\left[\log^2(1+x)/(1+x)\right](N)$, which can be calculated using
the representation (\ref{leq1b}) and lower order {\sc Mellin}
transforms. A simpler representation of $\Mvec\left
[\log^2(1-x)/(1+x)\right](N)$ is given by
%-----------------------------------------------------------------------
\begin{eqnarray}
\Mvec\left[\frac{\log^2(1-x)}{1+x}\right](N) &=&
~~\sum_{k=1}^9 a_k^{(1)} \Biggl\{
\frac{k}{N+k} \left[S_1^2(N+k) + S_2(N+k) \right] -
\left[S_1^2(k) + S_2(k) \right] \Biggr\} \nonumber\\ & &~~
+ \frac{7}{4} \zeta(3)
- \zeta(2) \log(2) + \frac{1}{3} \log^3(2)
\end{eqnarray}
%-----------------------------------------------------------------------
referring to (\ref{leq1}).
Similarly
%-----------------------------------------------------------------------
\begin{eqnarray}
\Mvec\left[\frac{\log(1-x)}{1+x}\right](N) &=&
- \sum_{k=1}^9 a_k^{(1)} \Biggl\{
\frac{k}{N+k} S_1(N+K) - S_1(k) \Biggr\} + \frac{1}{2} \left[ \log^2(2) -
\zeta(2) \right]~.
\end{eqnarray}
%-----------------------------------------------------------------------
holds. The above relations yield
%-----------------------------------------------------------------------
\begin{eqnarray}
\Mvec\left[\frac{\log(1-x) \Li_2(x)}{1+x}
\right](N)
       &=&
\sum_{k=0}^{11} c_k^{(1)} \left\{\frac{1}{2} \left[\log^2(2)
- \zeta(2) \right] \right. \nonumber\\ & & \left.
- \sum_{l=1}^9 a_l^{(1)} \left[\frac{l}{N+k+l}
S_1(N+k+l) - S_1(l) \right]\right\}
\nonumber\\ & &
+\sum_{k=0}^3 P_{2,k}^{(1)} \Biggl\{\frac{7}{4}\zeta(3)-\zeta(2)\log(2)
+\frac{1}{3} \log^3(2)         \nonumber\\ & &
+ \sum_{l=1}^9 a_l^{(1)} \Biggl[ \frac{l}{N+k+l}
\left[S_1^2(N+k+l)+S_2(N+k+l)\right] \nonumber\\  & &
-S_1^2(l)-S_2(l)\Biggr] \Biggr\}~.
\end{eqnarray}
%-----------------------------------------------------------------------

Finally, the {\sc Mellin}--transform of $\log(1+x) \Li_2(-x)/(1+x)$
is given by
%-----------------------------------------------------------------------
\begin{eqnarray}
\Mvec\left[\frac{\log(1+x) \Li_2(-x)}{1+x}
\right](N)
&=& - \frac{1}{4} \zeta(2) \log^2(2) 
+ \frac{1}{2} \left \{
\sum_{k=3}^{13} \frac{a_k^{(3)}}{N+k} \right.  \nonumber\\ & & \left.
+ \sum_{k=2}^{11} a_k^{(2)} 
\frac{N}{N+k} \left[\frac{1}{2} \zeta(2) - \frac{\log(2) - \beta(N+k+1)}
{N+k} \right]\right\}
\end{eqnarray}
%-----------------------------------------------------------------------
%%%%%%%%%%%%%%%%%%%%%%%%%%%%%%%%%%%%%%%%%%%%%%%%%%%%%%%%%%%%%%%%%%%%%%%
\section{Mellin Transforms for the
Functions
 $\mathbf{f(x)  /(x-1)} $}
\label{sec:mellinm}
%%%%%%%%%%%%%%%%%%%%%%%%%%%%%%%%%%%%%%%%%%%%%%%%%%%%%%%%%%%%%%%%%%%%%%%

\vspace{2mm}
\noindent
For this class of {\sc Mellin}--transforms the role of $\log(1+x)$ in
the previous section is taken by $\log(1-x)$, provided the functions
$f(x)$ in the numerator do vanish at a sufficient degree as $x \ra 1$.
Unlike $\log(1+x)$, the function $\log(1-x)$ has no simple polynomial
representation in the range $x~\epsilon~[0,1[$ at a comparable
accuracy to Eq.~(\ref{leq1}). One may be tempted to express the numerator
functions in a series of $\log(1-x)$ instead. This is indeed possible for
a wide class of {\sc Nielsen} integrals as ${\rm S}_{1,p}(x)$
in a simple 
manner, cf. also \cite{HV,DD},
%-----------------------------------------------------------------------
\begin{equation}
{\rm S}_{1,p}(x) = \frac{1}{p!} \sum_{k=0}^{\infty} \frac{B_k}{k!}
\frac{(-1)^k}{p+k} \log^{p+k}(1-x),
\end{equation}
%-----------------------------------------------------------------------
with $B_k$ the {\sc Bernoulli} numbers.\footnote{For other {\sc Nielsen}
integrals, as already for $\Li_l(x), l \geq 3$, the serial expansion in
$\log(1-x)$ leads no longer to optimal representations due to the
emergence of nested multiple series, see \cite{DD} for an example.}
The Taylor expansion of $\log^k(1-x)$ is given by
%-----------------------------------------------------------------------
\begin{eqnarray}
\log^k(1-x) =  (-1)^k k! \sum_{l=0}^{\infty} \left[\begin{array}{c}
n\\m\end{array}\right] \frac{x^l}{l!}~.
\label{logk}
\end{eqnarray}
%-----------------------------------------------------------------------
Here $\left[\begin{array}{c}
n\\m\end{array}\right]$ denote the {\sc Stirling}--numbers of the 
first kind~\cite{GKP}.
The {\sc Mellin}--transform is given by
%-----------------------------------------------------------------------
\begin{eqnarray}
\Mvec\left[\log^k(1-x)\right](N)
 =  (-1)^k k! \frac{1}{N} S_{\underbrace{\mbox{\scriptsize
1, \ldots, 1}}_k}(N)
 = \frac{\partial^k}{\partial b^k} \left.
 B(N  ,b+1)\right|_{b=0}~,
\end{eqnarray}
%-----------------------------------------------------------------------
where $B(a,b)$ denote     {\sc Euler}'s
Beta function.
The finite multiple harmonic sums $S_{\underbrace{\mbox{\scriptsize
1, \ldots, 1}}_k}(N)$
are given in \cite{BK2}, Eq.~(158),
and obey a determinant--representation.
The elements of the corresponding matrix are numbers and the single 
harmonic sums 
%-----------------------------------------------------------------------
\begin{eqnarray}
S_l(N) = \sum_{k=1}^N \frac{1}{k^l} = \frac{(-1)^{l-1}}{(l-1)!}
\psi^{(l-1)}(N+1) + c_l
\end{eqnarray}
%-----------------------------------------------------------------------
with $c_1 = \gamma_E$ and $c_l = \zeta(l)$ for $l > 1$. In this way
one obtains the analytic continuations. With growing values of $k$
the explicit expressions become rather lengthly. For numerical
computations one may use
the recursion relation, Eq.~(164), in Ref.~\cite{BK2},
%-----------------------------------------------------------------------
\begin{eqnarray}
S_{\underbrace{\mbox{\scriptsize
1, \ldots, 1}}_k}(N)= \frac{1}{k} \sum_{l=1}^k S_l(N)
S_{\underbrace{\mbox{\scriptsize
1, \ldots, 1}}_{k-l}}(N)
\end{eqnarray}
%-----------------------------------------------------------------------
for complex values of $N$. Alternatively to these exact expressions
one might also
use the Taylor series Eq.~(\ref{logk}) for which the {\sc
Mellin}--transform is easily calculated. The {\sc Stirling}--numbers of
the first kind
can be represented using a recursion relation,
cf.~Eqs.~(166,167) in Ref.~\cite{BK2}.
The {\sc Mellin} transform of ${\rm S}_{1,p}(x)$ is thus given by
%-----------------------------------------------------------------------
\begin{equation}
\label{s1pa}
\Mvec\left[{\rm S}_{1,p}(x)\right](N) = \frac{(-1)^p}{p!}
\sum_{k=0}^{\infty}
\frac{B_k}{k!} \frac{1}{p+k} \frac{1}{N}
S_{\underbrace{\mbox{\scriptsize
1, \ldots, 1}}_{p+k}}(N).
\end{equation}
%-----------------------------------------------------------------------
On the other hand the representation
%-----------------------------------------------------------------------
\begin{equation}
\label{s1pb}
\Mvec\left[{\rm S}_{1,p}(x)\right](N) = \frac{\zeta(n+1)}{N}
- \frac{1}{N^2}
S_{\underbrace{\mbox{\scriptsize
1, \ldots, 1}}_{p  }}(N)
\end{equation}
%-----------------------------------------------------------------------
is obtained by partial integration. This relation is more compact than 
the former. Eqs.~(\ref{s1pa},\ref{s1pb}) imply an interesting relation
between the multiple finite harmonic sums   of a single index and  the
{\sc Bernoulli} numbers and the $\zeta$--function.

Since not all {\sc Nielsen} integrals which we are considering in the
present paper can be expressed easily as a series in $\log(1-x)$ and
the corresponding series are found to converge not fast enough in a
series of cases we are going to apply a somewhat modified representation.
To guarantee a fast convergence also in the range of large values of
$x \lsim 1$ we subtract from some of the numerator functions a
polynomial in $\log(1-x)$ and $x$ of low degree. The remainder function
is expanded into a polynomial by the {\tt MINIMAX}--method. In both cases
the {\sc Mellin} transforms can be calculated analytically afterwards.
In some cases further partial integrations have to be performed.
Let us now discuss the individual cases in detail.

Some of the {\sc Mellin}--transforms are of the type
%-----------------------------------------------------------------------
\begin{equation}
\Mvec\left[\frac{\log^k(1+x) - \log^k(2)}{x-1} f(x)\right](N)~,
\label{Eqfaclog}
\end{equation}
%-----------------------------------------------------------------------
where $k=1,2$. The first factor in Eq.~(\ref{Eqfaclog}) can be
represented by a polynomial. 
Here the denominator is divided out of a
polynomial representation of the numerator. The corresponding
approximations are found applying the {\tt MINIMAX}--method,
%-----------------------------------------------------------------------
\begin{equation}
\frac{\log^l(1+x) - \log^l(2)}{x-1} \simeq \sum_{k=0}^{L(k)} b_k^{(l)}
x^k~,
\end{equation}
%-----------------------------------------------------------------------
with $L(1) = 8$ and $L(2) = 9$. The expansion coefficients are given
by
%-----------------------------------------------------------------------
\small
\begin{eqnarray}
%NEW
\begin{array}{lcrlcr}
b_0^{(1)}   &=&0.693147166991375\E+0~~~&
b_1^{(1)}   &=&-0.306850436868254\E+0
\nonumber\\
b_2^{(1)}   &=&0.193078041088284\E+0~~~&
b_3^{(1)}   &=&-0.139403892894644\E+0
\nonumber\\
b_4^{(1)}   &=&0.105269615988049\E+0~~~&
b_5^{(1)}    &=&-0.746801353858524\E-1
\nonumber\\
b_6^{(1)}  &=&0.427339135378207\E-1~~~&
b_7^{(1)}  &=&-0.161809049989783\E-1
\nonumber\\
b_8^{(1)}  &=&0.288664611077007\E-2~~~& & &
\end{array}
\nonumber
\end{eqnarray}
%-----------------------------------------------------------------------
\normalsize
and
\small
%-----------------------------------------------------------------------
\small
\begin{eqnarray}
%NEW
\begin{array}{lcrlcr}
b_0^{(2)}   &=&
                                  0.480453024731510\E+0~~~&
b_1^{(2)}   &=&
                                  0.480450679641120\E+0
\nonumber\\
b_2^{(2)}   &=&
                                  -0.519463586324817\E+0~~~&
b_3^{(2)}   &=&
                                  0.479285947990175\E+0
\nonumber\\
b_4^{(2)}   &=&
                                  -0.427765744446172\E+0~~~&
b_5^{(2)}    &=&
                                  0.360855321373065\E+0
\nonumber\\
b_6^{(2)}  &=&
                                  -0.263827078164263\E+0~~~&
b_7^{(2)}  &=&
                                  0.146927719341510\E+0
\nonumber\\
b_8^{(2)}  &=&
                                  -0.525105367350968\E-1~~~&
b_9^{(2)}  &=&
                                   0.874144396622167\E-2
\end{array}
\nonumber
\end{eqnarray}
\normalsize
%-----------------------------------------------------------------------
The related {\sc Mellin}--transforms have the representation
%-----------------------------------------------------------------------
\begin{eqnarray}
\Mvec\left[\frac{\log(1+x) - \log(2)}{x-1}\right](N)
 &=& \sum_{k=0}^8 \frac{b_k^{(1)}}{N+k+1}~,
\\
%------
\Mvec\left[\frac{\log^2(1+x) - \log^2(2)}{x-1}\right](N) 
 &=& \sum_{k=0}^9 \frac{b_k^{(2)}}{N+k+1}~,
\\
%------
\Mvec\left[\frac{\log(1+x) - \log(2)}{x-1} \Li_2(x)\right](N)
 &=& 
\sum_{k=0}^8 \frac{b_k^{(1)}}{N+k+1} \left[\zeta(2) -
\frac{S_1(N+k+1)}{N+k+1}\right]\\
%------
\Mvec\left[\frac{\log(1+x) - \log(2)}{x-1} \Li_2(-x)\right](N)
 &=& 
\sum_{k=0}^8 \frac{b_k^{(1)}}{N+k+1} \left[
- \frac{1}{2} \zeta(2) + \frac{\log(2) - \beta(N+k+2)}{N+k+1} \right]
\nonumber\\
\end{eqnarray}
%-----------------------------------------------------------------------

In the case of the remaining {\sc Mellin}--transforms of the type
%-----------------------------------------------------------------------
\begin{equation}
\Mvec\left[\frac{f(x)-f(1)}{x-1}\right](N)
\end{equation}
%-----------------------------------------------------------------------
   firstly the integral
%-----------------------------------------------------------------------
\begin{equation}
F(x) = \int_0^x dz
\frac{f(z)-f(1)}{z-1}
\end{equation}
%-----------------------------------------------------------------------
is evaluated leading to
%-----------------------------------------------------------------------
\begin{equation}
\Mvec\left[\frac{f(x)-f(1)}{x-1}\right](N)   = F(1) - N
\Mvec\left[F(x)\right
](N-1)~.
\end{equation}
%-----------------------------------------------------------------------
We now seek for an appropriate representation of the functions $F(x)$.
The respective functions are~:
%-----------------------------------------------------------------------
\begin{eqnarray}
\int_0^x dz \frac{\Li_2(z) - \zeta(2)}{z-1} &=& \left[\Li_2(x) -
\zeta(2)\right] \log(1-x) + 2 \Sf(x)
\\
%-------
\int_0^x dz \frac{\Li_2(-z) +
\zeta(2)/2}{z-1} &=& \left[\Li_2(-x) + \frac{1}{2} \zeta(2) \right] 
\log(1-x) + I_1(x)
\\
%-------
\int_0^x dz \frac{\Li_3(z) - \zeta(3)}{z-1} &=&  
\left[ \Li_3(x) - \zeta(3)\right] \log(1-x)
+ \frac{1}{2} \Li_2^2(x)
\\
%-------
\int_0^x dz \frac{\Sf(z) - \zeta(3)}{z-1} &=& \left[\Sf(x) - \zeta(3)
\right] \log(1-x) + 3 {\rm S}_{1,3}(x)
%-------
\end{eqnarray}
%-----------------------------------------------------------------------

Here one may use the representation
%-----------------------------------------------------------------------
\begin{eqnarray}
\Li_2(-x) = -\int_0^x \frac{dy}{y} \sum_{k=1}^9 a_k^{(1)} y^k
= -\sum_{k=1}^9 \frac{a_k^{(1)}}{k} x^k~.
\end{eqnarray}
%-----------------------------------------------------------------------
We derive approximations for $\Li_3(x), \Sf(x), \Li_2^2(x)$ and
$I_1(x)$ to be able to perform the {\sc Mellin}--transform using these
semi--analytic expressions. Here it is of particular importance to 
account for the $\log^k(1-x)$ behavior as $x \ra 1$.
One obtains
%-----------------------------------------------------------------------
\begin{eqnarray}
\Li_3(x) &\simeq&  P_0^{(2)}(x)+ \left[P_1^{(2)}(x)
+
\log(1-x) P_2^{(2)}(x)\right],
\\
\Sf(x)           &\simeq&  P_0^{(3)}(x)+ \left[P_1^{(3)}(x)
+\log(1-x) P_2^{(3)}(x)
+\log^2(1-x) P_3^{(3)}(x)
\right],
\\
\Li_2^2(x)       &\simeq&  P_0^{(4)}(x)+ \left[P_1^{(4)}(x)
+
\log(1-x) P_2^{(4)}(x)
+\log^2(1-x) P_3^{(4)}(x)\right]~,
\end{eqnarray}
%-----------------------------------------------------------------------
where $P_{0,1,2}^{(k)}(x)$ are the
polynomials  
%-----------------------------------------------------------------------
%-----------------------------------------------------------------------
\begin{eqnarray}
\label{eqP01}
P_0^{(2)}(x) &=& \sum_{k=0}^{12} c_k^{(2)} x^k\\
P_1^{(2)}(x) &=& \zeta(3) - \frac{11}{6} \zeta(2)
+ \frac{4}{3} + \left( 3 \zeta(2) - \frac{13}{4} \right) x
- \left(  \frac{3}{2} \zeta(2) - \frac{5}{2} \right)x^2
+ \left(\frac{1}{3} \zeta(2) - \frac{7}{12} \right) x^3
\nonumber\\ \\
P_2^{(2)}(x) &=&  -1 + \frac{5}{2} x - 2 x^2  + \frac{1}{2} x^3~,
\end{eqnarray}
%-----------------------------------------------------------------------
with
%-----------------------------------------------------------------------
\small
\begin{eqnarray}
%NEW
\begin{array}{lcrlcr}
c_0^{(2)} &=&                                  
0.480322239287459\E+0~~~&
c_1^{(2)} &=&                                  
-0.168480825099837\E+1
\nonumber\\
c_2^{(2)} &=&
0.209270571633447\E+1~~~&
c_3^{(2)} &=&
-0.101728150522737\E+1
\nonumber\\
c_4^{(2)} &=&
0.160180000661971\E+0~~~&
c_5^{(2)} &=&
-0.351982379713689\E+0
\nonumber\\
c_6^{(2)} &=&
0.141033369447519\E+1~~~&
c_7^{(2)} &=&
-0.353344124579927\E+1
\nonumber\\
c_8^{(2)} &=&
0.593934899678262\E+1~~~&
c_9^{(2)} &=&
-0.660019998525006\E+1
\nonumber\\
c_{10}^{(2)} &=&
0.466330491799074\E+1~~~&
c_{11}^{(2)} &=&
-0.189825521858848\E+1
\nonumber\\
c_{12}^{(2)} &=&
0.339773000152805\E+0~~~&
\nonumber\\
\end{array}
\nonumber
\end{eqnarray}
\normalsize
%-----------------------------------------------------------------------
%-----------------------------------------------------------------------
\begin{eqnarray}
\label{eqP02}
P_0^{(3)}(x) &=& \sum_{k=0}^{9}  c_k^{(3)} x^k\\
P_1^{(3)}(x) &=& \zeta(3) - \frac{2035}{1728} + \frac{205}{144} x
- \frac{95}{288} x^2 + \frac{43}{432} x^3 - \frac{1}{64} x^4\\
P_2^{(3)}(x) &=&  \frac{205}{144} - \frac{25}{12} x + \frac{23}{24} x^2
- \frac{13}{36} x^3 + \frac{1}{16} x^4 \\
P_3^{(3)}(x) &=&  -\frac{25}{24} +2 x - \frac{3}{2} x^2 + \frac{2}{3} x^3
- \frac{1}{8} x^4~,
\end{eqnarray}
%-----------------------------------------------------------------------
with
%-----------------------------------------------------------------------
\small
\begin{eqnarray}
%NEW
\begin{array}{lcrlcr}
c_0^{(3)} &=&
-0.243948949064443\E-1~~~&
c_1^{(3)} &=&
0.000005136294145\E+0
\nonumber\\
c_2^{(3)} &=&
0.249849075518710\E+0~~~&
c_3^{(3)} &=&
-0.498290708990997\E+0
\nonumber\\
c_4^{(3)} &=&
0.354866791547134\E+0~~~&
c_5^{(3)} &=&
-0.522116678353452\E-1
\nonumber\\
c_6^{(3)} &=&
-0.648354706049337\E-1~~~&
c_7^{(3)} &=&
0.644165053822532\E-1
\nonumber\\
c_8^{(3)} &=&
-0.394927322542075\E-1~~~&
c_9^{(3)} &=&
0.100879370657869\E-1
\nonumber\\
\end{array}
\end{eqnarray}
\normalsize
%-----------------------------------------------------------------------
%-----------------------------------------------------------------------
\begin{eqnarray}
\label{eqP03}
P_0^{(4)}(x) &=& \sum_{k=0}^{12}  c_k^{(4)} x^k\\
P_1^{(4)}(x) &=& 
\frac{257}{144} - \frac{205}{72} \zeta(2) +  \zeta(2)^2
- \left( \frac{167}{36} - \frac{25}{6} \zeta(2)\right) x
+ \left( \frac{101}{24} - \frac{23}{12} \zeta(2) \right) x^2
\nonumber\\ & &
- \left( \frac{59}{36} -  \frac{13}{18} \zeta(2) \right) x^3
+ \left( \frac{41}{144} - \frac{1}{8} \zeta(2) \right) x^4
\\
P_2^{(4)}(x) &=&  - \left(\frac{167}{36} - \frac{25}{6} \zeta(2)\right)
+ \left(\frac{235}{18} - 8 \zeta(2)
\right) x
- \left(\frac{40}{3} - 6 \zeta(2)\right) x^2
+ \left(\frac{109}{18} -\frac{8}{3} \zeta(2)\right) x^3
\nonumber\\ & &
- \left(\frac{41}{36} - \frac{1}{2} \zeta(2) \right) x^4 \\
P_3^{(4)}(x) &=&   \frac{35}{12} - \frac{26}{3}  x 
+ \frac{19}{2} x^2 -  \frac{14}{3} x^3  +
  \frac{11}{12} x^4~,
\end{eqnarray}
%-----------------------------------------------------------------------
with
%-----------------------------------------------------------------------
\small
\begin{eqnarray}
%NEW
\begin{array}{lcrlcr}
c_0^{(4)} &=&
0.192962504274437\E+0~~~&
c_1^{(4)} &=&
0.000005641557253\E+0
\nonumber\\
c_2^{(4)} &=&
-0.196891075399448\E+1~~~&
c_3^{(4)} &=&
0.392919138747074\E+1
\nonumber\\
c_4^{(4)} &=&
-0.290306105685546\E+1~~~&
c_5^{(4)} &=&
0.992890266001707\E+0
\nonumber\\
c_6^{(4)} &=&
-0.130026190226546\E+1~~~&
c_7^{(4)} &=&
0.341870577921103\E+1
\nonumber\\
c_8^{(4)} &=&
-0.576763902370864\E+1~~~&
c_9^{(4)} &=&
0.645554138192407\E+1
\nonumber\\
c_{10}^{(4)}&=&
0.459405622046138\E+1~~~&
c_{11}^{(4)}&=&
0.188510809558304\E+1
\nonumber\\
c_{12}^{(4)}&=&
-0.340476080290674\E+0~~~&
\end{array}
\end{eqnarray}
\normalsize
%-----------------------------------------------------------------------
These representations hold at an accuracy of better than $2 \times 
10^{-8}, 3 \times 10^{-8}$, and $2 \times 10^{-8}$, respectively,
for $x~\epsilon~[0,1]$.

In the case of the function $I_1(x)$, Eq.~(\ref{eqI1}),
the function $\log(1+x)$ in the 
integrand can be approximated by Eq.~(\ref{leq1}). 
The integral can then be
evaluated analytically and takes the form
%-----------------------------------------------------------------------
\begin{equation}
I_1(x) = P_0^{(5)}(x) + P_2^{(5)}(x) \log(1-x)~,
\end{equation}
%-----------------------------------------------------------------------
with
%-----------------------------------------------------------------------
\begin{eqnarray}
\label{eqP04}
P_0^{(5)}(x) &=& \sum_{k=1}^{9}  c_k^{(5)} x^k\\
P_2^{(5)}(x) &=& \sum_{k=0}^{9}  d_k^{(5)} x^k~.
\end{eqnarray}
%-----------------------------------------------------------------------
The coefficients $c_k^{(5)}$ and $d_k^{(5)}$ are algebraically
related to the coefficients $a_k^{(1)}$, Eq.~(\ref{leq1}). 
Their numerical values are
%-----------------------------------------------------------------------
\small
\begin{eqnarray}
\begin{array}{lcrlcr}
c_1^{(5)} &=&
-0.822467033400776\E+0~~~&
c_2^{(5)} &=&
 0.887664705657325\E-1
\nonumber\\
c_3^{(5)} &=&
-0.241549406045162\E-1~~~&
c_4^{(5)} &=&
 0.965074750946139\E-2
\nonumber\\
c_5^{(5)} &=&
-0.470587487919749\E-2~~~&
c_6^{(5)} &=&
 0.246014308378549\E-2
\nonumber\\
c_7^{(5)} &=&
-0.116431121874067\E-2~~~&
c_8^{(5)} &=&
 0.395705193848026\E-3 
\nonumber\\
c_9^{(5)} &=&
-0.664699010014505\E-4~~~& \\
\end{array}
\end{eqnarray}
%-----------------------------------------------------------------------
\begin{eqnarray}
\begin{array}{lcrlcr}
d_0^{(5)} &=&
-0.822467033400776\E+0~~~&
d_1^{(5)} &=&
 0.999999974532241\E+0
\nonumber\\
d_2^{(5)} &=&
-0.249997762945014\E+0~~~&
d_3^{(5)} &=&
 0.111067811851394\E+0
\nonumber\\
d_4^{(5)} &=&
-0.621323644338330\E-1~~~&
d_5^{(5)} &=&
 0.382902328987004\E-1
\nonumber\\
d_6^{(5)} &=&
-0.229110370338977\E-1~~~&
d_7^{(5)} &=&
 0.113158200819689\E-1
\nonumber\\
d_8^{(5)} &=&
-0.376387065979726\E-2~~~&
d_9^{(5)} &=&
 0.598229109013054\E-3
\end{array}
\end{eqnarray}
\normalsize
%-----------------------------------------------------------------------
This representation holds at
an accuracy of better than $3 \times 10^{-8}$
for $x~\epsilon~[0,1]$.

The {\sc Mellin}--transforms of the above polynomials are~:
%-----------------------------------------------------------------------
\begin{eqnarray}
\Mvec\left[\sum_{k=0}^m A_k x^k\right](N) &=& \sum_{k=1}^{m+1}
\frac{A_{k-1}}
{N+k} \\
\Mvec\left[\sum_{k=0}^{m} A_{k} x^k \log^l(1-x)\right](N)
&=& (-1)^l l! \sum_{k=1}^m \frac{A_{k-1}}{N+k}
S_{\underbrace{\mbox{\scriptsize 1, \ldots, 1}}_l}(N+k)~.
\end{eqnarray}
%-----------------------------------------------------------------------
Here,
%-----------------------------------------------------------------------
\begin{eqnarray}
S_{1,1}(N) &=& \frac{1}{2!} \left[S_1^2(N) + S_2(N)\right]\\
S_{1,1,1}(N) &=& \frac{1}{3!} \left[S_1^3(N) + 3 S_1(N) S_2(N)
+ 2 S_3(N) \right]~.
\end{eqnarray}
%-----------------------------------------------------------------------

The {\sc Mellin} transforms of the functions $g_{18}(x)-g_{22}(x)$
are then given represented by
%-----------------------------------------------------------------------
\begin{eqnarray}
\Mvec\left[\frac{\Li_2(x)-\zeta(2)}{x-1}\right](N)
&=& \frac{1}{N} \left[S_1^2(N) + S_2(N)\right] - \zeta(2) S_1(N)
+ \sum_{k=0}^{11} c_k^{(1)} \frac{N}{N+k} S_1(N+k) \nonumber\\ & &
-
\sum_{k=0}^3 P_{1,k}^{(2)}
 \frac{N}{N+k} \left[S_1^2(N+k) + S_2(N+k)\right]
\\
%-----------------------------------------------------------------------
\Mvec\left[\frac{\Li_2(-x)+\zeta(2)/2}{x-1}\right](N)
       &=& \frac{1}{2} \zeta(2) S_1(N) - \sum_{k=1}^9 \frac{a_k^{(1)}}{k}
S_1(N+k) \\
%-----------------------------------------------------------------------
\Mvec\left[\frac{\Li_3(x)-\zeta(3)}{x-1}\right](N)
       &=& \frac{1}{2} \zeta(2)^2 - \zeta(3) S_1(N) + \sum_{k=0}^{12}
c_k^{(2)} \frac{N}{N+k} S_1(N+k) \nonumber\\ & &
- \sum_{k=0}^3 P_{2,k}^{(2)} \frac{N}{N+k}
\left[S_1^2(N+k) + S_2(N+k)\right] - \frac{1}{2} \sum_{k=0}^{12}
c_k^{(4)} \frac{N}{N+k} \nonumber\\ & &
+ \frac{N}{2} \sum_{k=0}^4 \left[P_{2,k}^{(4)}
\frac{S_1(N+k)}{N+k} - P_{3,k}^{(4)} 
\frac{S_1^2(N+k) + S_2(N+k)}{N+k}\right]
\\
%-----------------------------------------------------------------------
\Mvec\left[\frac{\Sf(x)-\zeta(3)}{x-1}\right](N)
       &=& - \zeta(3) S_1(N) + \frac{1}{2 N}\left[S_1^3(N) + 2 S_1(N)
S_2(N) + 2 S_3(N) \right]  \nonumber\\ & &
+ \sum_{k=0}^9 c_k^{(3)} \frac{N}{N+k}
S_1(N+k)
+ \sum_{k=0}^4 \frac{N}{N+k} \left\{P_{3,k}^{(3)} \right. \nonumber\\
& & \left.
\left[
S_1^3(N+k) + 3 S_1(N+k) S_2(N+k) +2 S_3(N+k) \right]   \right.
\nonumber\\
        & &  \left.
-P_{2,k}^{(3)}
 \left[S_1^2(N+k) +S_2(N+k)\right]\right\}\\
%-----------------------------------------------------------------------
\Mvec\left[\frac{\log(x) \Li_2(x)}{x-1}\right](N)
       &=& \sum_{k=1}^{11} c_k^{(1)} \psi'(N+k+1) \nonumber\\ & &
- \sum_{k=0}^3
P_{2,k}^{(1)} \left[S_1(N+k) \psi'(N+k+1) - \frac{1}{2} \psi''(N+k+1)
\right]
\end{eqnarray}
%-----------------------------------------------------------------------

A set of other
integrals we are going to re--write in terms of integrals
of the type
$\int_0^x dz f_2(z)/(1+z)$, the {\sc Mellin}--transforms of
which are then evaluated using Eq.~(\ref{mxp1}). The following integral 
relations are obtained~:
%-----------------------------------------------------------------------
\begin{eqnarray}
%-------
\int_0^x dz \frac{\Li_3(-z)+(3/4)\zeta(3)
}{z-1} &=& \left[\Li_3(-x) 
+ \frac{3}{4} \zeta(3)\right]
\log(1-x) + \Li_2(x)
~\Li_2(-x)
\nonumber\\ & &
+ \Li_3(x) \log(1+x) - \int_0^x dz \frac{\Li_3(z)}{1+z} \\
%-------
\int_0^x dz \frac{I_1(z)+(5/8)\zeta(3)}{z-1} 
&=& \left[I_1(x) +\frac{5}{8} \zeta(3)\right]
\log(1-x) - 2 \Sf(x)
\log(1+x) \nonumber \\ & &
+ 2 \int_0^x dz \frac{\Sf(z)}{1+z}
\\
%-------
\int_0^x dz \frac{\Sf(-z)-\zeta(3)/8}{z-1} &=& \left[
\Sf(-x) -\frac{1}{8} \zeta(3)\right]
\log(1-x) - \frac{1}{2}
I_1(x) \log(1+x) \nonumber\\ & &
+ \frac{1}{2} \int_0^x dz \frac {I_1(z)}{1+z}
\end{eqnarray}
%-----------------------------------------------------------------------
Here one may represent $\Li_3(-x)$ and $\Sf(-x)$ by
%-----------------------------------------------------------------------
\begin{eqnarray}
\Li_3(-x) &=& - \sum_{k=1}^9 \frac{a_k^{(1)}}{k^2} x^k \\
\Sf(-x)   &=& \frac{1}{2} \sum_{k=2}^{11} \frac{a_k^{(2)}}{k} x^k~.
\end{eqnarray}
%-----------------------------------------------------------------------
The {\sc Mellin} transforms pf $g_{22}(x)-g_{25}(x)$ are represented
by
%-----------------------------------------------------------------------
\begin{eqnarray}
%-----------------------------------------------------------------------
\Mvec\left[\frac{\Li_3(-x)+3\zeta(3)/3}{x-1}\right](N)
&=& - \frac{1}{2} \zeta(2)^2 + \zeta(3) \log(2) + \frac{3}{4}
\zeta(3)
S_1(N) - \Mvec\left[ \frac{\Li_3(x)}{1+x}\right](N)
\nonumber\\ & & - \sum_{k=1}^9 a_k^{(1)}
\frac{N}{N+k} \left\{
\zeta(3) - \frac{\zeta(2)}{N+k} \right.
\nonumber\\
& &   \left.
- \frac{\zeta(2)}{k}
+ S_1(N+k) \left[ \frac{1}{(N+k)^2} + \frac{1}{k^2} + \frac{1}{k(N+k)}
\right]\right\}
\end{eqnarray}
%-----------------------------------------------------------------------
\begin{eqnarray}
\lefteqn{
\Mvec\left[\frac{I_1(x)+5\zeta(3)/8}{x-1}\right](N)=
- 2 \zeta(3) \log(2) + 2 \Mvec\left[\frac{S_{1,2}(x)}{1+x}
\right](N)}
    \nonumber\\
& &
+ \frac{5}{8} \zeta(3) S_1(N) 
+ \sum_{k=1}^9 a_k^{(1)} \frac{N}{N+k} \left[2 \zeta(3) 
- \frac{S_1^2(N+k)+S_2(N+k)}{N+k} \right] \nonumber\\ & &
+ \sum_{k=1}^9 c_k^{(5)}
\frac{N}{N+k} S_1(N+k)
- \sum_{k=0}^9 d_k^{(5)} \frac{N}{N+k} \left[ S_1^2(N+k) +S_2(N+k)
\right]
\end{eqnarray}
%-----------------------------------------------------------------------
\begin{eqnarray}
\Mvec\left[\frac{\Sf(-x)-\zeta(3)/8}{x-1}\right](N)
&=&  \frac{5}{16} \zeta(3) \log(2) + \frac{1}{2} \Mvec\left[ 
\frac{I_1(x)}{1+x}\right](N) - \frac{1}{8} \zeta(3) S_1(N)
\nonumber\\ & &
+ \frac{1}{2} \sum_{k=2}^{11} \frac{a_k^{(2)}}{k} \frac{N}{N+k}
S_1(N+k) + \frac{1}{2} \sum_{k=1}^9 a_k^{(1)} \frac{N}{N+k}
\left[- \frac{5}{8} \zeta(3) \right.
\nonumber\\ & &  \left.
+
 \sum_{l=1}^9 a_l^{(1)} \frac{S_1(N+k+l)}{N+k+l} \right]
\end{eqnarray}
%-----------------------------------------------------------------------

We finally would like to add a remark on a recent analysis of the
two--loop coefficient functions of deep inelastic scattering~\cite{MV}.
The representation given there also contains the harmonic sum
$S_{-1,1,1,1}(N)$ which was not needed to express the individual
{\sc Mellin} transforms in Refs.~\cite{NZ}, cf.~Ref.~\cite{BK2}, which 
may be caused due to the algebraic relations being applied in \cite{BK2}.
This sum has the representation
%-----------------------------------------------------------------------
\begin{eqnarray}
S_{-1,1,1,1}(N) = (-1)^{N+1} \frac{1}{6} \Mvec\left[
\frac{\log^3(1-x)}{1+x}\right](N) - \Li_4\left(\frac{1}{2}\right)~.
\end{eqnarray}
%-----------------------------------------------------------------------
The {\sc Mellin} transform of the function $\log^3(1-x)/(1+x)$ for
complex argument is easily obtained,
%-----------------------------------------------------------------------
\begin{eqnarray}
\Mvec\left[\frac{\log^3(1-x)}{1+x}\right](N) &=&
- \sum_{k=1}^9 a_k^{(1)} \Biggl\{
\frac{k}{N+k} \left[S_1^3(N+k) + 3 S_1(N+k)
S_2(N+k) + 2 S_3(N+k) \right]  \nonumber\\ 
& &~~~~~~~~~~~
- \left[S_1^3(k) + 3 S_1(k) S_2(k) + 2 S_3(k) \right] \Biggr\}
- 6 \Li_4\left(\frac{1}{2}\right)~.
\end{eqnarray}
%-----------------------------------------------------------------------

The above relations allow to express the {\sc Mellin} transforms of
all functions emerging in the coefficient functions and anomalous 
dimensions of massless gauge theories up to two--loop order and to
evaluate their analytic continuation from integer to complex arguments.
%%%%%%%%%%%%%%%%%%%%%%%%%%%%%%%%%%%%%%%%%%%%%%%%%%%%%%%%%%%%%%%%%%%%%%%
\section{The Code {\tt ANCONT}}
\label{sec:code}
%%%%%%%%%%%%%%%%%%%%%%%%%%%%%%%%%%%%%%%%%%%%%%%%%%%%%%%%%%%%%%%%%%%%%%%
%%%%%%%%%%%%%%%%%%%%%%%%%%%%%%%%%%%%%%%%%%%%%%%%%%%%%%%%%%%%%%%%%%%%%%%
\subsection{General Structure}
\label{sec:code4}
%%%%%%%%%%%%%%%%%%%%%%%%%%%%%%%%%%%%%%%%%%%%%%%%%%%%%%%%%%%%%%%%%%%%%%%

\vspace{2mm}
\noindent
The code {\tt ANCONT} calculates the {\sc Mellin}  transforms of the
basic functions $g_i(x)$ both for integer and complex values of the
{\sc Mellin} index N. The parameters of the code are initialized in
%-----------------------
\begin{center}
{\tt SUBROUTINE ACINI}.
\end{center}
%-----------------------
The calculations are performed in
%-----------------------
\begin{center}
{\tt SUBROUTINE ACRUN}. 
\end{center}
%-----------------------
The code ends with
%-----------------------
\begin{center}
{\tt SUBROUTINE ACEND}.
\end{center}
%-----------------------
%%%%%%%%%%%%%%%%%%%%%%%%%%%%%%%%%%%%%%%%%%%%%%%%%%%%%%%%%%%%%%%%%%%%%%%
\subsection{{\tt USER} routines}
\label{sec:code7}
%%%%%%%%%%%%%%%%%%%%%%%%%%%%%%%%%%%%%%%%%%%%%%%%%%%%%%%%%%%%%%%%%%%%%%%

\vspace{2mm}
\noindent
The user may access parts of the code via the user routines {\tt UINIT,
URUN} and {\tt UOUT}. These routines are called in {\tt ACINI, ACRUN}
and {\tt ACEND}, respectively. {\tt SUBROUTINE UINIT} may be used to
actualize the running parameters. Via {\tt URUN} the user may built
her/his own {\sc Mellin} convolutions or other functions of the
basic functions  and their {\sc Mellin} transforms for positive integer
or complex {\sc Mellin} index using the functions {\tt ACG$i$, FCT$i$}
or {\tt XCG$i$}, which are
described below, as a library. {\tt SUBROUTINE UOUT}
may be used as an output-interface at the end of the code.
To transfer data between different user routines the user may define
{\tt COMMON}-blocks named
\begin{center}
{\tt COMMON/USxxxx/ \ldots}
\end{center}
%%%%%%%%%%%%%%%%%%%%%%%%%%%%%%%%%%%%%%%%%%%%%%%%%%%%%%%%%%%%%%%%%%%%%%%
\subsection{Initialization}
\label{sec:code3}
%%%%%%%%%%%%%%%%%%%%%%%%%%%%%%%%%%%%%%%%%%%%%%%%%%%%%%%%%%%%%%%%%%%%%%%

\vspace{2mm}
\noindent
The main parameters and constants of the code are defined in
{\tt SUBROUTINE ACINI}. This routine calls the subroutines {\tt INVINI,
DEFAUL} and the user routine {\tt UINIT}. {\tt SUBROUTINE INVINI}
sets parameters and constants related to the {\sc Mellin} inversion.
The default running parameters  of the code are set in {\tt SUBROUTINE
DEFAUL}. These are~:

\vspace{3mm}
\noindent
\hspace*{2cm}
{\tt IRUN   = 0}\\
\hspace*{2cm}
{\tt ITEST1 = 1} \\
\hspace*{2cm}
{\tt ITEST2 = 1} \\
\hspace*{2cm}
{\tt ITEST3 = 1} \\
\hspace*{2cm}
{\tt IMIN   = 1} \\
\hspace*{2cm}
{\tt IMAX   = 26} \\
\hspace*{2cm}
{\tt IAPP   = 1} \\
\hspace*{2cm}
{\tt NMIN   = 1} \\
\hspace*{2cm}
{\tt NMAX   = 20} \\
\hspace*{2cm}
{\tt EPS    = 1.0D-9}

\vspace{3mm}
The running parameters are printed by {\tt SUBROUTINE WROUT}.
The parameters {\tt ITEST$i$} initialize tests of  the code, which are
inactive for {\tt ITEST$i$.NE.1}.
For {\tt ITEST1.EQ.1} the representation of the moments of the basic
functions $g_i(x)$ for positive
integer index by harmonic sums are compared to those obtained by
numerical integration. {\tt ITEST2.EQ.1} induces a comparison of the
representations for the analytic continuation of the {\sc Mellin}
transform of the basic functions with those by the harmonic sums
for positive integer index. For {\tt ITEST3.EQ.1} a comparison is made
for the {\sc Mellin} inversion using the analytic continuations of the
{\sc Mellin} transforms of the basic functions and the numerical
representations of the basic functions in the range $x~\epsilon~[10^{-7},
0.99]$.

{\tt IMIN} and {\tt IMAX} mark the index range of basic functions to
be used. Likewise {\tt NMIN} and {\tt NMAX} set the minimum and maximum
positive integer moment for the tests.

{\tt IAPP} selects the representation of the analytic continuation
for $g_3(x) = \Li_2(x)/(1+x)$ to compare the representations 
Eq.~(\ref{mxp2}) {\tt IAPP = 1} and
(\ref{eqgrv1}) {\tt IAPP=2}, (\ref{eqgrv2}) {\tt IAPP=3}, 
respectively.

{\tt EPS }~~{(\tt REAL*8)}
is a pilot parameter for the numerical integration of the
program {\tt DAIND}~\cite{AIND} and denotes the relative numerical
accuracy to be obtained.

The above parameters are available through the {\tt COMMON}-blocks

\vspace{3mm}
\noindent
\hspace*{2cm}
{\tt COMMON/TEST/ ITEST1, ITEST2,ITEST3}\\
\hspace*{2cm}
{\tt COMMON/RUN / IRUN }\\
\hspace*{2cm}
{\tt COMMON/MOMPA/ NMIN,NMAX}\\
\hspace*{2cm}
{\tt COMMON/FUNPA/ IMIN,IMAX}\\
\hspace*{2cm}
{\tt COMMON/IAPP/ IAPP}\\
\hspace*{2cm}
{\tt COMMON/EP/ EPS}.
%%%%%%%%%%%%%%%%%%%%%%%%%%%%%%%%%%%%%%%%%%%%%%%%%%%%%%%%%%%%%%%%%%%%%%%
\subsection{Running}
\label{sec:code2}
%%%%%%%%%%%%%%%%%%%%%%%%%%%%%%%%%%%%%%%%%%%%%%%%%%%%%%%%%%%%%%%%%%%%%%%

\vspace{2mm}
\noindent
The code provides three main lines, which are initialized setting
the parameters {\tt ITEST1, ITEST2, ITEST3 = 1}, respectively.

\noindent
For {\tt ITEST1= 1} the representations of the integer moments of the
basic functions $g_i(x)$ in terms of harmonic sums are tested comparing
them with the corresponding numerical integrals.
The value of the {\sc Mellin}
moment is calculated. The relative accuracy comparing both calculations
{\tt RAT = VAL1/VAL2 -1} and the value of the moment are provided.
The following test-output is obtained
for $k=9$, $g_9(x) = \Sf(-x)/(1+x)$, 
$n_1=2, n_2=20, val=1.0$D-9. Here {\tt N}
denotes the {\sc Mellin} index, {\tt RATk} the relative accuracy
comparing the numerical integration and the representation by
harmonic sums, and {\tt VAL} the value of the {\sc Mellin} moment.

{
\footnotesize
\begin{verbatim}
 ****************************************************************
 N,RAT9,VAL=  2   -1.983302411190380E-12   1.784970126521851E-02
 N,RAT9,VAL=  3   -1.992739306899693E-12   1.447619078834311E-02
 N,RAT9,VAL=  4   -1.742939126359033E-12   1.216425178651454E-02
 N,RAT9,VAL=  5   -1.722177955798543E-12   1.048399327332255E-02
 N,RAT9,VAL=  6   -1.469158128486470E-12   9.208975593568651E-03
 N,RAT9,VAL=  7   -1.485589429250922E-12   8.209008584718454E-03
 N,RAT9,VAL=  8   -1.187272502534142E-12   7.404084086538535E-03
 N,RAT9,VAL=  9   -1.256883486178140E-12   6.742389451876160E-03
 N,RAT9,VAL= 10   -1.036948304999896E-12   6.188923213501677E-03
 N,RAT9,VAL= 11   -8.926193117986259E-13   5.719203265306224E-03
 N,RAT9,VAL= 12   -8.733014311701481E-13   5.315598831318883E-03
 N,RAT9,VAL= 13   -7.922551503725117E-13   4.965094655720756E-03
 N,RAT9,VAL= 14   -5.925260282424460E-13   4.657875778975199E-03
 N,RAT9,VAL= 15   -6.212808045802376E-13   4.386402245697693E-03
 N,RAT9,VAL= 16   -3.986810881428937E-13   4.144786898528877E-03
 N,RAT9,VAL= 17   -4.581890422628021E-13   3.928366556207149E-03
 N,RAT9,VAL= 18   -1.886268918838141E-13   3.733399977739974E-03
 N,RAT9,VAL= 19   -3.899103262483550E-13   3.556850973662370E-03
 N,RAT9,VAL= 20   -1.719735465144367E-13   3.396229940556720E-03
 ****************************************************************
\end{verbatim}
}
\normalsize

\vspace{1mm}
\noindent
For {\tt ITEST2= 1} the representations of the {\sc Mellin} transforms of
the basic functions $g_i(x)$ valid for complex arguments
                    are compared to the
representation in terms of harmonic sums at positive integer argument.
The value of the {\sc Mellin}
moment is calculated. The relative accuracy comparing both calculations
{\tt RAT = VAL1/VAL2 -1} and the value of the moment are provided.
The relative accuracy of these representations are given in Table~2
for $N = 2$ to 20.
%%%%%%%%%%%%%%%%%%%%%%%%%%%%%%%%%%%%%%%%%%%%%%%%%%%%%%%%%%%%%%%%%%%%%%%%%
\renewcommand{\arraystretch}{1.3}
\begin{center}
\footnotesize
\begin{tabular}{||r||r|r|r|r|r|r||}
\hline \hline
\multicolumn{1}{||c}{$N$}&
\multicolumn{1}{|c}{ 1}&
\multicolumn{1}{|c}{ 2}&
\multicolumn{1}{|c}{ 3}&
\multicolumn{1}{|c}{ 4}&
\multicolumn{1}{|c}{ 5}&
\multicolumn{1}{|c||}{ 6}\\
\hline \hline
 2 & -1.06E-09 &  5.62E-10 & 7.61E-09 &
      3.52E-09 & -6.94E-09 & 4.67E-09 \\
%--------------------
 3 & -2.09E-09 &  1.06E-09 & 1.18E-08 &
      6.27E-09 & -1.15E-08 & 7.85E-09 \\
%--------------------
 4 & -3.48E-09 &  1.72E-09 & 1.69E-08 &
      9.92E-09 & -1.79E-08 & 1.20E-08 \\
%--------------------
 5 & -5.28E-09 &  2.54E-09 & 2.32E-08 &
      1.46E-08 & -2.58E-08 & 1.72E-08 \\
%--------------------
 6 & -7.50E-09 &  3.55E-09 & 3.07E-08 &
      2.04E-08 & -3.62E-08 & 2.36E-08 \\
%--------------------
 7 & -1.02E-08 &  3.97E-08 & 4.74E-09 &
      2.75E-08 & -4.92E-08 & 3.14E-08 \\
%--------------------
 8 & -1.34E-08 &  6.15E-09 & 5.01E-08 &
      3.61E-08 & -6.47E-08 & 4.07E-08 \\
%--------------------
 9 & -1.72E-08 &  7.78E-09 & 6.22E-08 &
      4.63E-08 & -8.21E-08 & 5.17E-08 \\
%--------------------
10 & -2.17E-08 &  9.67E-09 & 7.58E-08 &
      5.82E-08 & -1.01E-07 & 6.43E-08 \\
%--------------------
11 & -2.69E-08 &  1.18E-08 & 9.09E-08 &
      7.18E-08 & -1.12E-07 & 7.86E-08 \\
%--------------------
12 & -3.28E-08 &  1.43E-08 & 1.07E-07 &
      8.70E-08 & -1.38E-07 & 9.45E-08 \\
%--------------------
13 & -3.94E-08 &  1.70E-08 & 1.25E-07 &
      1.04E-07 & -1.56E-07 & 1.12E-07 \\
%--------------------
14 & -4.67E-08 &  2.01E-08 & 1.44E-07 &
      1.22E-07 & -1.72E-07 & 1.30E-07 \\
%--------------------
15 & -5.47E-08 &  2.34E-08 & 1.64E-07 &
      1.41E-07 & -1.85E-07 & 1.50E-07 \\
%--------------------
16 & -6.34E-08 &  2.71E-08 & 1.84E-07 &
      1.61E-07 & -1.96E-07 & 1.70E-07 \\
%--------------------
17 & -7.27E-08 &  3.11E-08 & 2.05E-07 &
      1.82E-07 & -2.03E-07 & 1.91E-07 \\
%--------------------
18 & -8.26E-08 &  3.54E-08 & 2.26E-07 &
      2.03E-07 & -2.08E-07 & 2.12E-07 \\
%--------------------
19 & -9.29E-08 &  3.99E-08 & 2.47E-07 &
      2.24E-07 & -2.09E-07 & 2.34E-07 \\
%--------------------
20 & -1.04E-07 &  4.47E-08 & 2.68E-07 &
      2.46E-07 & -2.07E-07 & 2.56E-07 \\
\hline \hline
\end{tabular}
\normalsize
\end{center}
%%%%%%%%%%%%%%%%%%%%%%%%%%%%%%%%%%%%%%%%%%%%%%%%%%%%%%%%%%%%%%%%%%%%%%%
\renewcommand{\arraystretch}{1}
\footnotesize
%%%%%%%%%%%%%%%%%%%%%%%%%%%%%%%%%%%%%%%%%%%%%%%%%%%%%%%%%%%%%%%%%%%%%%%%%
\renewcommand{\arraystretch}{1.3}
\begin{center}
\begin{tabular}{||r||r|r|r|r|r|r||}
\hline \hline
\multicolumn{1}{||c}{$N$}&
\multicolumn{1}{|c}{ 7}&
\multicolumn{1}{|c}{ 8}&
\multicolumn{1}{|c}{ 9}&
\multicolumn{1}{|c}{10}&
\multicolumn{1}{|c}{11}&
\multicolumn{1}{|c||}{12}\\
\hline \hline
%****************************************
2 & 
 3.67\E-09 &
 2.47\E-08 &
 6.04\E-09 &
-9.99\E-09 &
 9.43\E-09 &
-3.16\E-10 \\
%----------------------------------------
3 &
 6.48\E-09 &
 3.25\E-08 &
 9.77\E-09 &
-1.03\E-08 &
 1.40\E-08 &
-6.46\E-10 \\
%----------------------------------------
4 &
 1.02\E-08 &
 4.15\E-08 &
 1.45\E-08 &
-9.44\E-09 &
 1.93\E-08 &
-1.05\E-09 \\
%----------------------------------------
5 &
 1.49\E-08 &
 5.16\E-08 &
 2.02\E-08 &
-7.45\E-09 &
 2.54\E-08 &
 -1.52\E-09 \\
%----------------------------------------
6 &
 2.08\E-08 &
 6.29\E-08 &
 2.72\E-08 &
-4.20\E-09 &
 3.22\E-08 &
-2.07\E-09 \\
%----------------------------------------
7 &
 2.80\E-08 &
 7.56\E-08 &
 3.57\E-08 &
 4.18\E-10 &
 3.98\E-08 &
-2.72\E-09 \\
%----------------------------------------
8 &
3.67\E-08 &
8.94\E-08 &
4.57\E-08 &
6.48\E-09 &
4.80\E-08 &
-3.46\E-09 \\
%----------------------------------------
9 &
4.71\E-08 &
1.04\E-07 &
5.74\E-08 &
1.40\E-08 &
5.68\E-08 &
-4.31\E-09\\
%----------------------------------------
10 &
5.91\E-08 &
1.20\E-07 &
7.09\E-08 &
2.31\E-08 &
6.59\E-08 &
-5.28\E-09\\
%----------------------------------------
11 &
7.28\E-08 &
1.37\E-07 &
8.60\E-08 &
3.36\E-08 &
7.53\E-08 &
-6.37\E-09\\
%----------------------------------------
12 &
8.81\E-08 &
1.54\E-07 &
1.03\E-07 &
4.55\E-08 &
8.48\E-08 &
-7.57\E-09\\
%----------------------------------------
13 &
1.05\E-07 &
1.72\E-07 &
1.21\E-07 &
5.90\E-08 &
9.42\E-08 &
-8.89E-09\\
%----------------------------------------
14 &
1.23\E-07 &
1.91\E-07 &
1.40\E-07 &
7.28\E-08 &
1.03\E-07 &
-1.03\E-08\\
%----------------------------------------
15 & 
1.42\E-07 &
2.09\E-07 &
1.60\E-07 &
8.80\E-08 &
1.12\E-07 &
-1.18\E-08\\
%----------------------------------------
16 &
1.62\E-07 &
2.27\E-07 &
1.81\E-07 &
1.04\E-07 &
1.21\E-07 &
-1.34\E-08\\
%----------------------------------------
17 &
1.83\E-07 &
2.46\E-07 &
2.03\E-07 &
1.21\E-07 &
1.29E-07  &
-1.51\E-08\\
%----------------------------------------
18 &
2.04\E-07 &
2.64\E-07 &
2.26\E-07 &
1.38\E-07 &
 1.36E-07 &
-1.68\E-08\\
%----------------------------------------
19 &
2.26\E-07 &
2.81\E-07 &
2.48\E-07 &
1.55\E-07 &
1.43\E-07 &
-1.86\E-08\\
%----------------------------------------
20 & 
2.48\E-07 &
2.98\E-07 &
2.71\E-07 &
1.73\E-07 &
1.49\E-07 &
-2.04\E-08 \\
\hline \hline
\end{tabular}
\end{center}

\vspace{2mm}
\normalsize
\noindent
{\sf Table~2: Relative accuracy of the {\sc Mellin} transforms
comparing the approximative representations valid for complex arguments
and the representation in terms of harmonic sums for $N~\epsilon~[2,20]$
}
%%%%%%%%%%%%%%%%%%%%%%%%%%%%%%%%%%%%%%%%%%%%%%%%%%%%%%%%%%%%%%%%%%%%%%%
\renewcommand{\arraystretch}{1}
%%%%%%%%%%%%%%%%%%%%%%%%%%%%%%%%%%%%%%%%%%%%%%%%%%%%%%%%%%%%%%%%%%%%%%%%%
\renewcommand{\arraystretch}{1.3}
\footnotesize
\begin{center}
\begin{tabular}{||r||r|r|r|r|r|r||}
\hline \hline
\multicolumn{1}{||c}{$N$}&
\multicolumn{1}{|c}{13}&
\multicolumn{1}{|c}{14}&
\multicolumn{1}{|c}{15}&
\multicolumn{1}{|c}{16}&
\multicolumn{1}{|c}{17}&
\multicolumn{1}{|c||}{18}\\
\hline \hline
%****************************************
%-----------------------------------------------
2 &
-9.79\E-10 &
-2.70\E-10 &
-9.02\E-10 &
-7.02\E-10 &
 9.31\E-10 &
 -6.00\E-10 \\
%----------------------------------------
3 &
-2.08\E-09 &
-3.63\E-10 &
-1.14\E-09 &
-9.13\E-10 &
 1.18\E-09 &
 -1.13E-09 \\
%----------------------------------------
4 &
-3.56\E-09 &
-4.59\E-10 &
-1.40\E-09 &
-1.14\E-09 &
 1.44\E-09 &
 -1.80E-09 \\
%----------------------------------------
5 &
-5.45\E-09 &
-5.60\E-10 &
-1.68\E-09 &
-1.34\E-09 &
 1.72\E-09 &
 -2.60E-09 \\
%----------------------------------------
6 &
-7.80\E-09 &
-6.65\E-10 &
-1.96\E-09 &
-1.64\E-09 &
 2.01\E-09 &
 -3.54E-09 \\
%----------------------------------------
7 &
-1.06\E-08 &
-7.75\E-10 &
-2.27\E-09 &
-1.92\E-09 &
 2.32\E-09 &
 -4.60E-09 \\
%----------------------------------------
8 & 
-1.40\E-08 &
-8.93\E-10 &
-2.60\E-09 &
-2.23\E-09 &
 2.65\E-09 &
 -5.80E-09 \\
%----------------------------------------
9 & 
-1.81\E-08 &
-1.02\E-09 &
-2.95\E-09 &
-2.57\E-09 &
 3.00\E-09 &
 -7.14E-09 \\
%----------------------------------------
10 & 
-2.28\E-08 &
-1.15\E-09 &
-3.32\E-09 &
-2.92\E-09 &
 3.39\E-09 &
 -8.60E-09 \\
%----------------------------------------
11 &
-2.82\E-08 &
-1.30\E-09 &
-3.69\E-09 &
-3.29\E-09 &
 3.79\E-09 &
 -1.02E-08 \\
%----------------------------------------
12 &
-3.43\E-08 &
-1.45\E-09 &
-4.06\E-09 &
-3.68\E-09 &
 4.22\E-09 &
 -1.19E-08 \\
%----------------------------------------
13 &
-4.11\E-08 &
-1.62\E-09 &
-4.43\E-09 &
-4.06\E-09 &
 4.66\E-09 &
 -1.38E-08 \\
%----------------------------------------
14 &
-4.86\E-08 & 
-1.78\E-09 &
-4.79\E-09 &
-4.44\E-09 &
 5.11\E-09 &
 -1.58E-08 \\
%----------------------------------------
15 &
-5.68\E-08 &
-1.96\E-09 &
-5.14\E-09 &
-4.80\E-09 &
 5.56\E-09 &
 -1.80E-08 \\
%----------------------------------------
16 &
-6.55\E-08 &
-2.14\E-09 &
-5.47\E-09 &
-5.16\E-09 &
 6.02\E-09 &
 -2.04E-08 \\
%----------------------------------------
17 &
-7.49\E-08 &
-2.31\E-09 &
-5.77\E-09 &
-5.49\E-09 &
 6.47\E-09 &
 -2.29E-08 \\
%----------------------------------------
18 &
-8.47\E-08 &
-2.49\E-09 &
-6.06\E-09 &
-5.80\E-09 &
 6.91\E-09 &
 -2.56E-08 \\
%----------------------------------------
19 &
-9.49\E-08 &
-2.66\E-09 &
-6.32\E-09 &
-6.09\E-09 &
 7.34\E-09 &
 -2.85E-08 \\
%----------------------------------------
20 &
-1.06\E-07 &
-2.83\E-09 & 
-6.55\E-09 & 
-6.36\E-09 &
 7.75\E-09 &
 -3.15E-08 \\
\hline \hline
\end{tabular}
\end{center}

\vspace{2mm}
\noindent
\normalsize
%%%%%%%%%%%%%%%%%%%%%%%%%%%%%%%%%%%%%%%%%%%%%%%%%%%%%%%%%%%%%%%%%%%%%%%
\footnotesize
\renewcommand{\arraystretch}{1}
%%%%%%%%%%%%%%%%%%%%%%%%%%%%%%%%%%%%%%%%%%%%%%%%%%%%%%%%%%%%%%%%%%%%%%%
%%%%%%%%%%%%%%%%%%%%%%%%%%%%%%%%%%%%%%%%%%%%%%%%%%%%%%%%%%%%%%%%%%%%%%%
\renewcommand{\arraystretch}{1}
%%%%%%%%%%%%%%%%%%%%%%%%%%%%%%%%%%%%%%%%%%%%%%%%%%%%%%%%%%%%%%%%%%%%%%%%%
\renewcommand{\arraystretch}{1.3}
\begin{center}
\begin{tabular}{||r||r|r|r|r|r|r|r|r||}
\hline \hline
\multicolumn{1}{||c}{$N$}&
\multicolumn{1}{|c}{19}&
\multicolumn{1}{|c}{20}&
\multicolumn{1}{|c}{21}&
\multicolumn{1}{|c}{22}&
\multicolumn{1}{|c}{23}&
\multicolumn{1}{|c}{24}&
\multicolumn{1}{|c}{25}&
\multicolumn{1}{|c||}{26}\\
\hline \hline
%****************************************
%-----------------------------------------------
2 &
 -0.248\E-08 &
  0.108\E-08 &
  0.646\E-09 &
 -0.131\E-09 &
  0.431\E-08 &
 -0.408\E-08 &
  0.353\E-08 &
 -0.489\E-08 \\
%-----------------------------------------------
3 &
 -0.303\E-08 &
  0.212\E-08 &
  0.114\E-08 &
 -0.166\E-09 &
  0.686\E-08 &
 -0.500\E-08 &
  0.461\E-08 &
 -0.795\E-08 \\
%-----------------------------------------------
4 &
 -0.353\E-08 &
  0.351\E-08 &
  0.174\E-08 &
 -0.200\E-09 &
  0.963\E-08 &
 -0.586\E-08 &
  0.568\E-08 &
 -0.115\E-07 \\
%-----------------------------------------------
5 &
 -0.397\E-08 &
  0.525\E-08 &
  0.243\E-08 &
 -0.235\E-09 &
  0.126\E-07 &
 -0.668\E-08 &
  0.674\E-08 &
 -0.154\E-07 \\
%-----------------------------------------------
6 &
 -0.435\E-08 &
  0.733\E-08 &
  0.321\E-08 &
 -0.270\E-09 &
  0.157\E-07 &
 -0.746\E-08 &
  0.779\E-08 &
 -0.198\E-07 \\
%-----------------------------------------------
7 &
 -0.470\E-08 &
  0.976\E-08 &
  0.409\E-08 &
 -0.305\E-09 &
  0.189\E-07 &
 -0.821\E-08 &
  0.882\E-08 &
 -0.248\E-07 \\
%-----------------------------------------------
8 &
 -0.499\E-08 &
  0.126\E-07 &
  0.506\E-08 &
 -0.341\E-09 &
  0.223\E-07 &
 -0.893\E-08 &
  0.985\E-08 &
 -0.303\E-07 \\
%-----------------------------------------------
9 &
 -0.525\E-08 &
  0.157\E-07 &
 -0.377\E-09 &
  0.612\E-08 &
  0.257\E-07 &
 -0.963\E-08 &
  0.109\E-07 &
 -0.365\E-07 \\
%-----------------------------------------------
10 &
 -0.546\E-08 &
  0.192\E-07 &
  0.728\E-08 &
 -0.413\E-09 &
  0.293\E-07 &
 -0.103\E-07 &
  0.118\E-07 &
 -0.432\E-07 \\
%-----------------------------------------------
11 &
 -0.562\E-08 &
  0.231\E-07 &
  0.855\E-08 &
 -0.449\E-09 &
  0.329\E-07 &
 -0.110\E-07 &
  0.128\E-07 &
 -0.506\E-07 \\
%-----------------------------------------------
12 &
 -0.575\E-08 &
  0.274\E-07 &
  0.993\E-08 &
 -0.486\E-09 &
  0.366\E-07 &
 -0.116\E-07 &
  0.138\E-07 &
 -0.587\E-07 \\
%-----------------------------------------------
13 &
 -0.584\E-08 &
  0.321\E-07 &
  0.114\E-07 &
 -0.523\E-09 &
  0.404\E-07 &
 -0.123\E-07 &
  0.147\E-07 &
 -0.675\E-07 \\
%-----------------------------------------------
14 &
 -0.590\E-08 &
  0.371\E-07 &
  0.131\E-07 &
 -0.561\E-09 &
  0.442\E-07 &
 -0.129\E-07 &
  0.157\E-07 &
 -0.770\E-07 \\
%-----------------------------------------------
15 &
 -0.592\E-08 &
  0.426\E-07 &
  0.148\E-07 &
 -0.600\E-09 &
  0.481\E-07 &
 -0.135\E-07 &
  0.166\E-07 &
 -0.872\E-07 \\
%-----------------------------------------------
16 &
 -0.592\E-08 &
  0.485\E-07 &
  0.166\E-07 &
 -0.641\E-09 &
  0.520\E-07 &
 -0.141\E-07 &
  0.175\E-07 &
 -0.981\E-07 \\
%-----------------------------------------------
17 &
 -0.589\E-08 &
  0.549\E-07 &
  0.186\E-07 &
 -0.683\E-09 &
  0.560\E-07 &
 -0.147\E-07 &
  0.184\E-07 &
 -0.110\E-06 \\
%-----------------------------------------------
18 &
 -0.583\E-08 &
  0.617\E-07 &
  0.207\E-07 &
 -0.729\E-09 &
  0.601\E-07 &
 -0.153\E-07 &
  0.192\E-07 &
 -0.122\E-06 \\
%-----------------------------------------------
19 &
 -0.576\E-08 &
  0.690\E-07 &
  0.229\E-07 &
 -0.776\E-09 &
  0.642\E-07 &
 -0.159\E-07 &
  0.201\E-07 &
 -0.135\E-06 \\
%-----------------------------------------------
20 &
 -0.566\E-08 &
  0.769\E-07 &
  0.251\E-07 &
 -0.827\E-09 &
  0.683\E-07 &
 -0.165\E-07 &
  0.210\E-07 &
 -0.149\E-06 \\
%-----------------------------------------------
\hline \hline
\end{tabular}
\end{center}

\vspace{2mm}
\normalsize
\noindent
{\sf Table~2~:~continued}
%%%%%%%%%%%%%%%%%%%%%%%%%%%%%%%%%%%%%%%%%%%%%%%%%%%%%%%%%%%%%%%%%%%%%%%
\renewcommand{\arraystretch}{1}
%%%%%%%%%%%%%%%%%%%%%%%%%%%%%%%%%%%%%%%%%%%%%%%%%%%%%%%%%%%%%%%%%%%%%%%

\noindent
For {\tt ITEST3= 1} the {\sc Mellin} inversion for the basic functions
$g_i(x)$ based on the representations of the {\sc Mellin} transforms
for complex argument are compared to the numerical expressions
for the functions $g_i(x)$ in the range $x~\epsilon~[10^{-7},~0.99]$.
The value of $g_i$ as a function of $x$
is calculated. The relative accuracy comparing both calculations
{\tt RAT = VAL1/VAL2 -1} is given.
The inverse {\sc Mellin} transform to $x$-space
is performed
by a
contour integral numerically. The singularities of the {\sc Mellin}
transforms for ${\rm N}~\epsilon~{\bf C}$ are situated at the 
non--positive integers for the coefficient and splitting functions in
massless fixed--order perturbation theory.~\footnote{Some all order 
resummations, as e.g. the small--$x$ resummation, lead also to
singularities outside the real axis, cf.~\cite{LOWX}.} The 
non--perturbative input densities are usually expressed in terms of
polynomials of the type
%-----------------------------------------------------------------------
\begin{eqnarray}
h(x) = \sum_i A_i x^{\alpha_i} (1-x)^{\beta_i}~,
\end{eqnarray}
%-----------------------------------------------------------------------
the {\sc Mellin} transform of which is a sum of {\sc Euler} 
Beta--functions. Their singularities are as well situated on the real
axis left of an upper bound. The inverse {\sc Mellin} transform
is given by
%-----------------------------------------------------------------------
\begin{eqnarray}
h(x) = \int_0^{\infty} dz {\sf Im} \left[ e^{i\Phi} x^{-c(z)} f[c(z)]
\right]
\end{eqnarray}
%-----------------------------------------------------------------------
where $c(z) = c_0 + z e^{i\Phi}$, cf. also~\cite{GRV}.
The parameter integral over $z$ can be performed by standard algorithms.
The code segments the integration path logarithmically into {\tt N=20}
or {\tt N=50} pieces on each of which the integral is calculated by the
8-point or 32-point {\sc Gauss} formula. The starting point {\tt C}
is varied according to the rightmost singularity of the function
to be inverted and is chosen close to it on the right. The angle of the
linear path w.r.t. the real axis is chosen by $\Phi = (3/4) \pi$ for
$x < 0.8$, $\Phi = (7/8) \pi$ for $x > 0.8$.
and  $\Phi = (19/20) \pi$ for $x > 0.98$.

As an example the relative
accuracy of the representation for the function
$g_{16} = \left[\log(1+x) - \log(2)\right]/(x-1) \Li_2(-x)$ between
$x=10^{-7}$ and $x=0.99$ is

{
\footnotesize
\begin{verbatim}
******************************************************
X,RAT16,VAL=  .10000D-06    -.44220D-08    -.69315D-07
X,RAT16,VAL=  .10000D-05    -.10120D-07    -.69315D-06
X,RAT16,VAL=  .10000D-04    -.13908D-07    -.69314D-05
X,RAT16,VAL=  .10000D-03    -.16350D-07    -.69310D-04
X,RAT16,VAL=  .10000D-02    -.15094D-07    -.69267D-03
X,RAT16,VAL=  .10000D-01     .61713D-08    -.68838D-02
X,RAT16,VAL=  .50000D-01     .94884D-08    -.33499D-01
X,RAT16,VAL=  .10000D+00    -.19742D-07    -.64836D-01
X,RAT16,VAL=  .15000D+00    -.92017D-08    -.94219D-01
X,RAT16,VAL=  .20000D+00     .14400D-07    -.12183D+00
X,RAT16,VAL=  .25000D+00     .20299D-07    -.14783D+00
X,RAT16,VAL=  .30000D+00     .50955D-08    -.17236D+00
X,RAT16,VAL=  .35000D+00    -.15289D-07    -.19554D+00
X,RAT16,VAL=  .40000D+00    -.23563D-07    -.21747D+00
X,RAT16,VAL=  .45000D+00    -.13795D-07    -.23827D+00
X,RAT16,VAL=  .50000D+00     .61787D-08    -.25800D+00
X,RAT16,VAL=  .55000D+00     .21289D-07    -.27676D+00
X,RAT16,VAL=  .60000D+00     .19772D-07    -.29461D+00
X,RAT16,VAL=  .65000D+00     .16678D-08    -.31161D+00
X,RAT16,VAL=  .70000D+00    -.19671D-07    -.32783D+00
X,RAT16,VAL=  .75000D+00     .22255D-05    -.34332D+00
X,RAT16,VAL=  .80000D+00     .22936D-05    -.35811D+00
X,RAT16,VAL=  .85000D+00     .30694D-05    -.37226D+00
X,RAT16,VAL=  .90000D+00     .38631D-05    -.38581D+00
X,RAT16,VAL=  .95000D+00    -.25871D-02    -.39879D+00
X,RAT16,VAL=  .99000D+00    -.20484D-01    -.40879D+00
******************************************************
\end{verbatim}
\normalsize
}
Similar accuracies are obtained also in the other cases.
We would like to mention that for $i = 3$ and {\tt IAPP=2,3}, 
respectively the accuracy is lower by three orders of magnitude up to
$x \sim 0.6$ relative to the results for {\tt IAPP=1}
and remains less above.

\vspace{2mm}
\noindent
The central functions of the code
are~:
                                {\tt FCT$i$(N), XCG$i$, ACG$i$} and
{\tt FKN$i$}.
The {\tt REAL*8 FUNCTIONS} {\tt FCT$i$(N),~i=1..26} provide the
positive integer moments of the basic functions $g_i(x)$, 
{\tt N $\geq$ 1}
obtained by numerical integration using the program 
{\tt DAIND}~\cite{AIND}. The integrands are defined in the
{\tt REAL*8 FUNCTIONS} {\tt FKT$i$(N),~i=1..26}.

A second representation of the positive integer moments of the basic
functions is given using the finite harmonic sums in the
{\tt REAL*8 FUNCTIONS} {\tt XCG$i$(N),~i=1..26}.

The analytic continuations of the {\sc Mellin} transforms are given in
the 
{\tt COMPLEX*16 FUNCTIONS} {\tt ACG$i$(N),~i=1..26}.

The basic functions $g_i(x)$ are given by 
{\tt REAL*8 FUNCTION FKN$i$(X),~i=1..26}.

\vspace{1mm}
One may use the above options without performing the aforementioned
comparisons. The corresponding pilot parameter is {\tt IRUN}, which
takes values between {\tt 0} and {\tt 5}.\\

\vspace{1mm}
\noindent
{\tt IRUN = 0:}~~implies a test run.

\vspace{1mm}
\noindent
{\tt IRUN = 1:}~~The {\sc Mellin} moments of the basic functions are
calculated by numerical integration for positive {\sc Mellin} indices.

\vspace{1mm}
\noindent
{\tt IRUN = 2:}~~The {\sc Mellin} moments of the basic functions are
calculated by their representation in terms of harmonic sums for
positive {\sc Mellin} indices.

\vspace{1mm}
\noindent
{\tt IRUN = 3:}~~The {\sc Mellin} moments of the basic functions are
calculated by their representation in terms of the analytic continuations
{\tt ACG$i$} for complex argument at  positive integer moments.

\vspace{1mm}
\noindent
{\tt IRUN = 4:}~~The basic functions $g_i(x)$
are calculated for $x~\epsilon~[10^{-7},0.99]$ using numerical
representations for the corresponding functions including the
{\sc Nielsen} integrals.

\vspace{1mm}
\noindent
{\tt IRUN = 5:}~~The basic functions $g_i(x)$
are calculated for $x~\epsilon~[10^{-7},0.99]$ using the {\sc Mellin}
moment inversion by a complex contour integral.
%%%%%%%%%%%%%%%%%%%%%%%%%%%%%%%%%%%%%%%%%%%%%%%%%%%%%%%%%%%%%%%%%%%%%%%
\subsection{Subsidiary Routines}
\label{sec:code1}
%%%%%%%%%%%%%%%%%%%%%%%%%%%%%%%%%%%%%%%%%%%%%%%%%%%%%%%%%%%%%%%%%%%%%%%

\vspace{2mm}
\noindent
A series of subsidiary routines representing mathematical functions
and procedures are contained in the code.

%%%%%%%%%%%%%%%%%%%%%%%%%%%%%%%%%%%%%%%%%%%%%%%%%%%%%%%%%%%%%%%%%%%%%%%
\subsubsection{$\mathbf{\psi^{(k)}(z)}$}
%%%%%%%%%%%%%%%%%%%%%%%%%%%%%%%%%%%%%%%%%%%%%%%%%%%%%%%%%%%%%%%%%%%%%%%

\vspace{1mm}
\noindent
\begin{center}
{\tt SUBROUTINE PSI$k$(ZZ,RES),~~k = 0..3}
\end{center}
provides the value of the functions {\tt RES = $\psi^{(k)}($ZZ$)$}.
Both {\tt ZZ} and {\tt RES} are {\tt COMPLEX*16}.
%%%%%%%%%%%%%%%%%%%%%%%%%%%%%%%%%%%%%%%%%%%%%%%%%%%%%%%%%%%%%%%%%%%%%%%
\subsubsection{$\mathbf{\beta^{(k)}(z)}$}
%%%%%%%%%%%%%%%%%%%%%%%%%%%%%%%%%%%%%%%%%%%%%%%%%%%%%%%%%%%%%%%%%%%%%%%

\vspace{1mm}
\noindent
\begin{center}
{\tt SUBROUTINE BET$k$(ZZ,RES),~~k = 0..3}
\end{center}
provides the value of the functions {\tt RES = $\beta^{(k)}($ZZ$)$},
cf.~Eq.~(\ref{eqbeta}).
Both {\tt ZZ} and {\tt RES} are
{\tt COMPLEX*16}.
%%%%%%%%%%%%%%%%%%%%%%%%%%%%%%%%%%%%%%%%%%%%%%%%%%%%%%%%%%%%%%%%%%%%%%%
\subsubsection{$\mathbf{\Gamma(z)}$}
%%%%%%%%%%%%%%%%%%%%%%%%%%%%%%%%%%%%%%%%%%%%%%%%%%%%%%%%%%%%%%%%%%%%%%%

\vspace{1mm}
\noindent
\begin{center}
{\tt SUBROUTINE GAMMA(Z,RES)}
\end{center}
is the {\sc Euler} Gamma function {\tt RES = $\Gamma($Z$)$} for complex
argument, with
%-----------------------------------------------------------------------
\begin{equation}
\Gamma(z+1) = z \Gamma(z)~.
\end{equation}
%-----------------------------------------------------------------------
{\tt Z} and {\tt RES} are {\tt COMPLEX*16}. 
For $|z| > 10, {\rm |arg|}(z) <
\pi$ the function is represented by
%-----------------------------------------------------------------------
\begin{eqnarray}
\log[\Gamma(z)] &\simeq&
z \left[\log(z) - 1\right] + \frac{1}{2} \log(2\pi/z)
\\
& & + \frac{1}{12} \frac{1}{z} - \frac{1}{360} \frac{1}{z^3}
+ \frac{1}{1260} \frac{1}{z^5} - \frac{1}{1680} \frac{1}{z^7}
- \frac{691}{360360} \frac{1}{z^{11}} + \frac{1}{156} \frac{1}{z^{13}}
- \frac{3617}{122400} \frac{1}{z^{15}}~. \nonumber
\end{eqnarray}
%-----------------------------------------------------------------------
The {\tt SUBROUTINE GAMMAL(Z,RES)} delivers
{\tt RES = $\log[\Gamma($Z$)]$}.
%%%%%%%%%%%%%%%%%%%%%%%%%%%%%%%%%%%%%%%%%%%%%%%%%%%%%%%%%%%%%%%%%%%%%%%
\subsubsection{$\mathbf{B(a,b)}$}
%%%%%%%%%%%%%%%%%%%%%%%%%%%%%%%%%%%%%%%%%%%%%%%%%%%%%%%%%%%%%%%%%%%%%%%

\vspace{1mm}
\noindent
\begin{center}
{\tt SUBROUTINE BETA(A,B,RES)}
\end{center}
is the {\sc Euler} Beta function {\tt RES = $B($A,B$)$} for complex
argument, with
%-----------------------------------------------------------------------
\begin{eqnarray}
B(a,b) = 
\frac{\Gamma(a) \Gamma(b)}{\Gamma(a+b)}~.
\end{eqnarray}
%-----------------------------------------------------------------------
{\tt A,B} and {\tt RES} are {\tt COMPLEX*16}.
This function may be used to represent the {\sc Mellin} transforms
of the partonic input densities for complex argument.
%%%%%%%%%%%%%%%%%%%%%%%%%%%%%%%%%%%%%%%%%%%%%%%%%%%%%%%%%%%%%%%%%%%%%%%
\subsubsection{Harmonic Sums}
%%%%%%%%%%%%%%%%%%%%%%%%%%%%%%%%%%%%%%%%%%%%%%%%%%%%%%%%%%%%%%%%%%%%%%%

\vspace{1mm}
\noindent
The code contains routines for the (alternating) harmonic sums up to
depth 5 as {\tt REAL*8 FUNCTIONS}:
\begin{center}
{\tt DOUBLE PRECISION FUNCTION SUM1(I1,N)} \\
%$\cdot$\\ $\cdot$\\
$\vdots$ \\
{\tt DOUBLE PRECISION FUNCTION SUM5(I1,I2,I3,I4,I5,N)}
\end{center}
The input parameters     {\tt I$i$} may be either positive
(non--alternating summation) or negative (alternating summation) integers.

\noindent
{\sf Example~:~} $S_{1,-1,3}(N)$ is calculated by {\tt SUM3(1,-1,3,N)}.
%%%%%%%%%%%%%%%%%%%%%%%%%%%%%%%%%%%%%%%%%%%%%%%%%%%%%%%%%%%%%%%%%%%%%%%
\subsubsection{Polylogarithms}
%%%%%%%%%%%%%%%%%%%%%%%%%%%%%%%%%%%%%%%%%%%%%%%%%%%%%%%%%%%%%%%%%%%%%%%

\vspace{1mm}
\noindent
The {\tt REAL*8 FUNCTIONS}
\begin{center}
{\tt DOUBLE PRECISION FUNCTION FLI$i$(x),~~i=2,3,4}
\end{center}
calculate the polylogarithms $\Li_k(x),~k=2,3,4$ for $x~\epsilon[-1,1]$.
%%%%%%%%%%%%%%%%%%%%%%%%%%%%%%%%%%%%%%%%%%%%%%%%%%%%%%%%%%%%%%%%%%%%%%%
\subsubsection{Nielsen Integral $\mathbf{S_{12}(x)}$}
%%%%%%%%%%%%%%%%%%%%%%%%%%%%%%%%%%%%%%%%%%%%%%%%%%%%%%%%%%%%%%%%%%%%%%%

\vspace{1mm}
\noindent
The Nielsen integral $S_{1,2}(x)$, cf.~Eq.~(\ref{eqsnp}), is calculated
by
\begin{center}
{\tt DOUBLE PRECISION FUNCTION S12$(x)$}.
\end{center}
%%%%%%%%%%%%%%%%%%%%%%%%%%%%%%%%%%%%%%%%%%%%%%%%%%%%%%%%%%%%%%%%%%%%%%%
\subsubsection{$\mathbf{I_1(x)}$}
%%%%%%%%%%%%%%%%%%%%%%%%%%%%%%%%%%%%%%%%%%%%%%%%%%%%%%%%%%%%%%%%%%%%%%%

\vspace{1mm}

\noindent
The function $I_1(x)$, Eq.~(\ref{eqI1}) is provided by
\begin{center}
{\tt DOUBLE PRECISION FUNCTION YI1$(x)$}.
\end{center}
%%%%%%%%%%%%%%%%%%%%%%%%%%%%%%%%%%%%%%%%%%%%%%%%%%%%%%%%%%%%%%%%%%%%%%%
\subsubsection{Numerical Integration}
%%%%%%%%%%%%%%%%%%%%%%%%%%%%%%%%%%%%%%%%%%%%%%%%%%%%%%%%%%%%%%%%%%%%%%%

\vspace{1mm}
\noindent
\begin{center}
{\tt REAL*8 FUNCTION DAIND(A,B,FUN,EPS,KEY,MAX,KOU,EST)}
\end{center}
The function {\tt DAIND} yields the integral over the function
{\tt FUN(x)} from {\tt A} to {\tt B}. The integrand has to be declared as
{\tt EXTERNAL} in the calling routine.
The corresponding algorithm was published in
Ref.~\cite{AIND}. We recommend to use the integrator setting
{\tt KEY = 2}. Here {\tt EPS} denotes the demanded relative accuracy of 
the integral and {\tt MAX $\leq$ 10000} the number of points at which
{\tt FUN(x)} is calculated by this adaptive integration. The output
parameters {\tt KOU} and {\tt EST} refer to the number of points being
used and the estimated accuracy reached.
%%%%%%%%%%%%%%%%%%%%%%%%%%%%%%%%%%%%%%%%%%%%%%%%%%%%%%%%%%%%%%%%%%%%%%%
\section{Summary}
\label{sec:conc}
%%%%%%%%%%%%%%%%%%%%%%%%%%%%%%%%%%%%%%%%%%%%%%%%%%%%%%%%%%%%%%%%%%%%%%%

\vspace{2mm}
\noindent
We have calculated semi-analytic representations for the analytic
continuations of the {\sc Mellin} transforms of the set of basic
functions through which the {\sc Wilson} coefficients and splitting
functions which occur in hard scattering processes in massless field
theories as QED and QCD can be expressed up to two--loop order.
Here we aimed on high--precision representations which were performed
using widely the algebraic and analytic relations of the {\sc Nielsen}
integrals and related functions being considered as well as their
{\sc Mellin}
transforms limiting the necessary approximative representations to as
few cases as possible. The expressions obtained are compared both
to the {\sc Mellin} moments at positive integer {\sc Mellin} index
with the values obtained by numerical integration and the
representations in terms of harmonic sums.
  The {\sc Mellin} inversion
to $x$ space provides a further test on the numerical accuracy of
the expressions derived for the analytic continuations of the
{\sc Mellin} transforms for complex values.
The {\tt FORTRAN}--code
{\tt ANCONT} is provided.
With the help of these representations the {\sc Mellin} transforms
for all two--loop quantities for the different      space-- and
time--like hard scattering processes in the massless limit
can be assembled.

\vspace{2mm}
\noindent
{\b Acknowledgement}.~This work was supported in part by EU contract
FMRX-CT98-0194 (DG 12 - MIHT).
%%%%%%%%%%%%%%%%%%%%%%%%%%%%%%%%%%%%%%%%%%%%%%%%%%%%%%%%%%%%%%%%%%%%%%%%%
%%%%%%%%%%%%%%%%%%%%%%%%%%%%%%%%%%%%%%%%%%%%%%%%%%%%%%%%%%%%%%%%%%%%%%%

%%%%%%%%%%%%%%%%%%%%%%%%%%%%%%%%%%%%%%%%%%%%%%%%%%%%%%%%%%%%%%%%%%%%%%%%%
\end{document}